\documentclass[reqno,10pt,a4paper]{amsart}
\usepackage[margin=2.7cm]{geometry}
\usepackage{cite}
\usepackage{newpxtext}
\usepackage{cabin}
\usepackage[varqu,varl]{inconsolata}
\usepackage[bigdelims,vvarbb]{newpxmath}
\usepackage[cal=cm,bb=esstix,bbscaled=1.05]{mathalfa}
\usepackage{graphicx}
\usepackage{booktabs}
\usepackage{hyperref}
\usepackage{longtable}
\usepackage{subcaption}
\usepackage{tikz-cd}
\usepackage{wrapfig}
\usepackage[ruled,vlined]{algorithm2e}
\usepackage{tabularx}
\usepackage{multirow}
\usepackage{bm}

\hypersetup{
 colorlinks,
 linkcolor={red!50!black},
 citecolor={blue!50!black},
 urlcolor={blue!80!black}
}
\newcommand{\orcid}[1]{\,\resizebox{8px}{!}{\href{https://orcid.org/#1}{\includegraphics{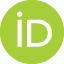}}}}
\graphicspath{{images/}}
\newcommand{\ZZ}{\mathbb{Z}}
\newcommand{\QQ}{\mathbb{Q}}
\newcommand{\FF}{\mathbb{F}}
\newcommand{\doi}[1]{\href{https://doi.org/#1}{doi:#1}}

\theoremstyle{definition}
\newtheorem*{Ex*}{Example}
\begin{document}
\author[Y.-H.\ He]{Yang-Hui He\orcid{0000-0002-0787-8380}}
\address{London Institute for Mathematical Sciences, Royal Institution, London, W1S 4BS, UK\newline
\indent Department of Mathematics, City St George's, University of London, London, EC1V 0HB, UK\newline
\indent Merton College, University of Oxford, Oxford, OX1 4JD, UK
}
\email{yh@lims.ac.uk}
\author[A.\ M.\ Kasprzyk]{Alexander M.~Kasprzyk\orcid{0000-0003-2340-5257}}
\address{Mathematics Institute, University of Warwick, Coventry, CV4 7AL, UK}
\email{alexander.kasprzyk@warwick.ac.uk}
\author{Q Le\orcid{0000-0002-8172-2464}}
\address{Mathematical Institute, University of Oxford, Oxford, OX2 6GG, UK}
\email{q.le@trinity.ox.ac.uk}
\author[D.\ Riabchenko]{Dmitrii Riabchenko\orcid{0009-0005-8099-8628}}
\address{Department of Mathematics, City St George's, University of London, London, EC1V 0HB, UK}
\email{Dmitrii.Riabchenko@citystgeorges.ac.uk}
\title{Machine Learning Discovers New Champion Codes}
\maketitle
\begin{abstract}
Linear error-correcting codes form the mathematical backbone of modern digital communication and storage systems, but identifying champion linear codes (linear codes achieving or exceeding the best known minimum Hamming distance) remains challenging. By training a transformer to predict the minimum Hamming distance of a class of linear codes and pairing it with a genetic algorithm over the search space, we develop a novel method for discovering champion codes. This model effectively reduces the search space of linear codes needed to achieve champion codes. Our results present the use of this method in the study and construction of error-correcting codes, applicable to codes such as generalised toric, Reed--Muller, Bose--Chaudhuri--Hocquenghem, algebrogeometric, and potentially quantum codes.
\end{abstract}
\section{Introduction}
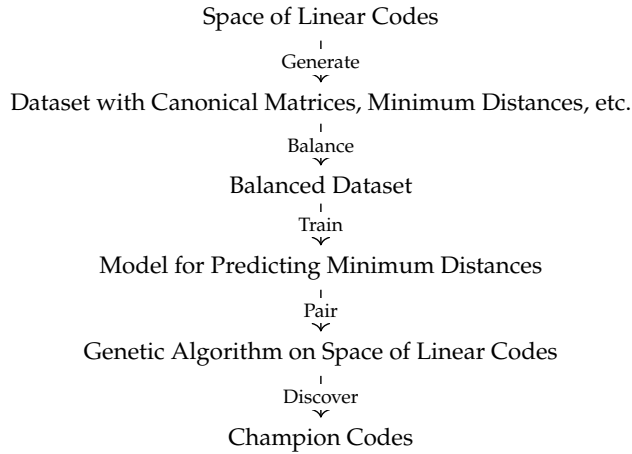
\begin{figure}[h]
 \centering
 \small
 \begin{tikzcd}
	{\text{Space of Linear Codes}} & {} \\
	{\text{Dataset with Canonical Matrices, Minimum Distances, etc.}} \\
	{\text{Balanced Dataset}} \\
	{\text{Model for Predicting Minimum Distances}} \\
	{\text{Genetic Algorithm on Space of Linear Codes}} \\
	{\text{Champion Codes}}
	\arrow["{\text{Generate}}"{description}, from=1-1, to=2-1]
	\arrow["{\text{Balance}}"{description}, from=2-1, to=3-1]
	\arrow["{\text{Train}}"{description}, from=3-1, to=4-1]
	\arrow["{\text{Pair}}"{description}, from=4-1, to=5-1]
	\arrow["{\text{Discover}}"{description}, from=5-1, to=6-1]
\end{tikzcd}
 \caption{A general method for the discovery of champion error-correcting codes.}
 \label{fig:the paradigm}
\end{figure}

Coding theory forms the foundation of all modern communication. Error-correcting codes -- used to detect and correct errors in data transmitted over Wi-Fi and LAN networks~\cite{kadel2020opportunities}, under transatlantic waters~\cite{pamart1994forward}, and across deep space~\cite{fredricksen1968error,heller1971viterbi} -- are a key part of coding theory, and the search for more efficient error-correcting codes is of great importance. A crucial property when determining an error-correcting code's capabilities is its minimum Hamming distance. Finding new linear codes whose minimum Hamming distance equals or exceeds that of previously known codes (for fixed block length and dimension) is computationally challenging, and is an NP-hard problem~\cite{vardy1997intractability}; these codes are called~\emph{champion codes}. We present a new method for discovering champion linear codes, employing machine learning and genetic algorithms, as summarised in Figure~\ref{fig:the paradigm}.

To demonstrate the efficacy of this method, we restrict ourselves to a class of error-correcting linear codes called generalised toric codes, introduced by Hansen~\cite{jhansen1} and Little~\cite{little1}, however there are many other classes of codes that could be used~\cite{reed1960polynomial, gallager2003low, arikan2009channel}. We develop a transformer architecture, adapted from a simplified version of OpenAI's GPT-2 model~\cite{radford2019language}, to train a predictive model for the minimum Hamming distance~$d$ of a generalised toric code over~$\FF_7$, achieving $\sim$91.6\% accuracy with a tolerance of $d \pm 3$ and mean absolute error 1.05 on the test set. Combining this model with a genetic algorithm, we rediscover champion codes initially found in~\cite{kasprzyk1}. Using a similar method over~$\FF_8$ we find over 500 champion codes, of which at least six are new (see Table~\ref{tab:182238examplecodes}). A comparison of our method with a random search suggests that it achieves up to a twofold improvement in~BZ evaluations, significantly reducing the computational cost of discovering new champion codes. A~BZ evaluation is a use of the Brouwer--Zimmermann algorithm, the brute-force algorithm for determining the minimum Hamming distance of any linear code.

Our method builds upon previous work by Brown--Kasprzyk~\cite{BrownKasprzyk13,kasprzyk1} which systematically classified all generalised toric codes over~$\FF_q$, $q\leq 7$. The methods used there failed to extend to~$\FF_8$ due to the exponential computational cost of enumerating all candidate codes and calculating the minimum Hamming distance. Our method is, in principle, more general, and can be applied to any family of linear codes with an evolvable parameter space (see the section on `Evolving champion codes' below).

This paper follows a tradition of applying the techniques of data science and machine learning to purely mathematical data. This tradition has its origins in the study of string theory and algebraic geometry~\cite{he2017machine}, and has expanded to other areas of mathematics; for example~\cite{DaviesEtAl2021,WuDeLoera2022,CoatesKasprzykVeneziale23,CoatesKasprzykVeneziale24,he2024ai,Gukov:2024buj,yau2024graph,berglund2024new,Abramson2024,hashemi2025can}. Though machine learning does not guarantee optimal solutions to all problems, our investigation shows a successful application of machine learning to an NP-hard problem~\cite{ardon2022reinforcement} and a new methodology for discovering champion codes.

\section{Methods}
\subsection*{Generalised toric codes}
We focus on generalised toric codes~\cite{little1}, a class of linear codes which have proven useful for discovering new champion codes~\cite{diego1,little1,BrownKasprzyk13,kasprzyk1}. See~\S\ref{app:sec1} of the Appendix for the basics of coding theory, as well as a summary of the standard notation.

\begin{figure}[t]
\centering
\small
\begin{tabular}{@{}c@{}c@{}c@{}c@{}c@{}}
$V$&&$C_V(\FF_5)$&&$d$\\
\begin{tikzpicture}[scale=0.65]
 \foreach \y in {0,1,2,3} 
 \draw[thick] (0,\y) -- (3,\y);
 \foreach \x in {0,1,2,3} 
 \draw[thick] (\x,0) -- (\x,3);
 \fill[black] (2,0) circle (3pt);
 \fill[black] (0,1) circle (3pt);
 \fill[black] (1,1) circle (3pt);
 \fill[black] (3,1) circle (3pt);
 \fill[black] (1,2) circle (3pt);
 \fill[black] (2,3) circle (3pt);
\end{tikzpicture}
&
\raisebox{1cm}{
$\longrightarrow$
}
&
\raisebox{1cm}{
$\left(\begin{array}{@{\ }c@{\ \ }c@{\ \ }c@{\ \ }c@{\ \ }c@{\ \ }c@{\ \ }c@{\ \ }c@{\ \ }c@{\ \ }c@{\ \ }c@{\ \ }c@{\ \ }c@{\ \ }c@{\ \ }c@{\ \ }c@{\ }}
1&0&0&0&0&1&3&0&0&2&0&4&4&2&2&1\\
0&1&0&0&0&0&1&3&0&2&1&3&2&1&1&0\\
0&0&1&0&0&4&3&2&0&4&0&2&0&2&1&1\\
0&0&0&1&0&1&1&2&0&1&1&4&1&0&2&1\\
0&0&0&0&1&2&4&3&0&0&0&0&4&3&1&2\\
0&0&0&0&0&0&0&0&1&2&4&3&1&2&4&3
\end{array}\right)$
}
&
\raisebox{1cm}{
$\longrightarrow$
}
&
\raisebox{1cm}{
$8$
}
\end{tabular}
 \caption{An example of a generalised toric code~$C_V$ over~$\FF_5$, with $V = \{(0,1),(1,1), (1,2),(2,0), (2,3),(3,1)\}$ and minimum Hamming distance eight.}
 \label{fig:example_toriccode}
\end{figure}
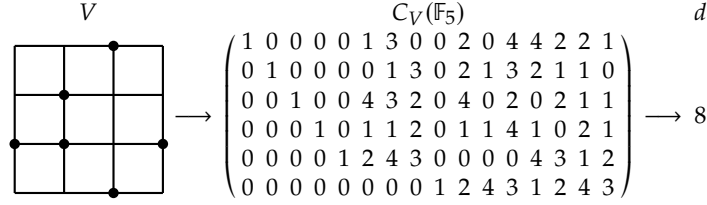

Fix a finite field~$\FF_q$, for some prime power~$q$, and primitive element~$\xi \in \FF_q^*$, and consider the planar lattice grid~$[0,q-2]^2$. For each~$u = (a,b) \in [0,q-2]^2$, define the map 
\begin{align*}
e_u \colon \FF_q^* \times \FF_q^* & \longrightarrow \FF_q \\
(\xi^i, \xi^j) & \longmapsto (\xi^i)^{a} (\xi^j)^{b}
\end{align*}
Let~$V = \{(a_1,b_1), \dots, (a_m,b_m)\}$ be a set of $m$ lattice points in~$[0,q-2]^2$. The~\emph{generalised toric code} $C_V(\FF_q)$ over the field~$\FF_q$, associated to the lattice points~$V$, is the linear code spanned by the vectors
\begin{equation*}
\left\{ (e_u((\xi^i,\xi^j)) : 0 \leq i,j \leq q-2) : u \in V \right\}
 =
\left\{ (\xi^{ia + jb} : 0 \leq i,j \leq q-2) : (a,b) \in V \right\}
\end{equation*}
This has block length~$n=(q-1)^2$. An example of this construction is illustrated in Figure~\ref{fig:example_toriccode}. When~$q$ is specified, we omit the field~$\FF_q$ from the notation.

If we restrict to (non-generalised) toric codes~\cite{hansen2000toric,jhansen1}, we can related them to divisors on toric varieties. In particular, results on cohomology and intersection theory for toric varieties can be used to obtain bounds on the minimum Hamming distance~\cite{joyner,minkowskilength,jhansen1}. Although more is known in the context of (non-generalised) toric codes, we consider the generalised case for several reasons: constructing datasets for generalised toric codes is computationally cheaper; they have been successfully used to discover champion codes~\cite{kasprzyk1}; their simple parametrisation makes them suitable for genetic optimisation (see the section on `Genetic algorithms' below); and they provide a non-trivial test case for our method.

We conclude this section by summarising the method we employ to find champion codes:
\begin{enumerate}
 \item consider a space of linear codes that have an evolvable parameter space, as explained in the section on `Evolving champion codes' below;
 \item generate a dataset of each linear code's canonical generator matrix (or, equivalently, parity check matrix) and minimum Hamming distance;
 \item balance this dataset to allow for training and testing datasets;
 \item train a model to predict the minimum distance of a code with this balanced dataset;
 \item pair this model with a genetic algorithm that evolves over the space of linear codes;
 \item run the genetic algorithm several times and discover new error-correcting codes, possibly finding champion codes.
\end{enumerate}

\subsection*{Machine learning the minimum Hamming distance}
The problem of computing the minimum Hamming distance of a code is an NP-hard problem and is typically calculated using the Brouwer--Zimmermann~(BZ) algorithm~\cite{zimmermann1996integral, grassl2006searching}. Despite several improvements, the~BZ algorithm remains computationally costly, having polynomial in~$k$ and exponential in~$q$ complexity~\cite{hernando2019algorithm}. Thus our method seeks to minimise the number of~BZ evaluations required to identify new champion codes (see Tables~\ref{tab:efficiencyF7} and~\ref{tab:efficiencyF8} for results).

A linear code~$C$ can be represented by a $k\times n$ generator matrix~$G$ (or, equivalently, a parity check matrix~$P$) whose rows form a basis of~$C$; here $n$ is the block length of $C$ and $k$ is the dimension of $C$. Thus the problem can be reformulated as follows: given an input sequence~$S$ of~$k$ codewords, each of length~$n$, predict the minimum Hamming distance $d$ (a non-negative integer bounded above by $n$). This task belongs to a family of problems called~\emph{sequence modelling}, for which several architectures are commonly used: convolutional neural networks, recurrent neural networks, and transformers. The latter two naturally handle variable-length input sequences (although transformers impose a maximum sequence length).

Since computing a code’s minimum Hamming distance is NP-hard, we employ the most powerful sequence modelling architecture currently available, the transformer~\cite{vaswani2017attention}, which underlies many state-of-the-art natural language processing~(NLP) models such as ChatGPT~\cite{brown2020language} and sentiment analysis systems~\cite{barbieri2020tweeteval}. The latter task is conceptually analogous to ours: given a sequence of elements (text tokens or, analogously, codewords), the goal is to predict a class label (text sentiment or, analogously, minimum Hamming distance). Accordingly, we sometimes refer to codes of a given minimum distance as belonging to a~\emph{class}.
 
To train this model, we constructed a dataset consisting of triples~$(V,G,d)$, where $V$ is a set of points in~$[0,q-2]^2$ defining a generalised toric code~$C_V(\FF_q)$, $G$~is the canonical generator matrix of~$C_V$, and~$d$ is the minimum Hamming distance of~$C_V$. See~\cite{GAs} for an implementation using \texttt{SageMath}~\cite{sagemath}, \texttt{Magma}~\cite{BosmaCannonPlayoust1997}, and~\texttt{PyTorch}~\cite{pytorch}. Further details of the models can be found in~\S\ref{sec:appendixML} of the Appendix.

\subsection*{Datasets of generalised toric codes}
\begin{figure}[t]
\centering
\includegraphics[width=1\textwidth]{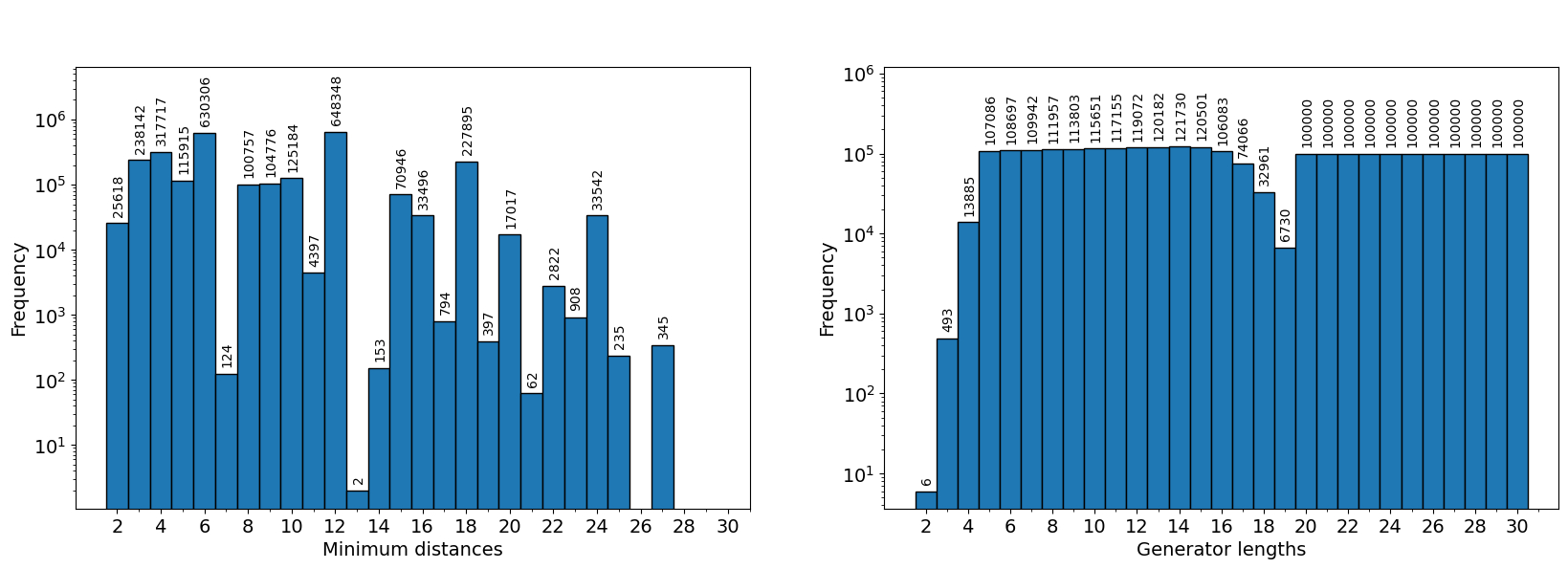}
\caption{Histograms of the~$\FF_7$ codes dataset. The left histogram shows the distribution of the minimum Hamming distances. The right histogram shows the distribution of the generators' dimensions.}
\label{fig:F7data}
\end{figure}

The dataset generated for generalised toric codes over~$\FF_7$ contains 2,700,000 elements. This data is summarised in Figure~\ref{fig:F7data}. Notice that there is significant variation in the frequencies of the minimum Hamming distances, ranging from just two cases to more than 600,000. As the dataset is unbalanced, splitting the data into train and test sets randomly would be inadequate. Instead, we split subsets with equal minimum distance into train and test sets separately and merged them into full train and test sets with oversampling to 500 or downsampling to 50,000 if necessary, and re-shuffling afterwards. A resulting dataset of approximately 550,000 elements was used for training with 9:1 train/test split.

\begin{figure}[t]
\centering
\includegraphics[width=1\textwidth]{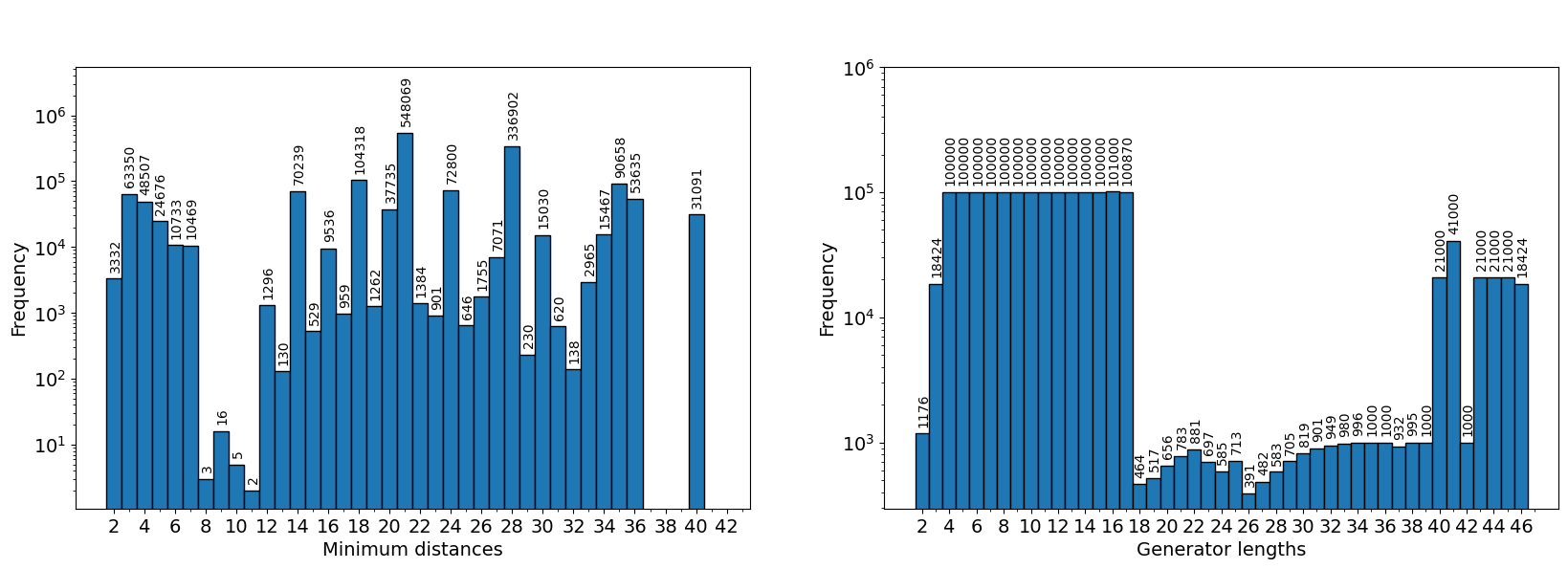}
\caption{Histograms of the~$\FF_8$ codes dataset. The left histogram shows the distribution of the minimum Hamming distances. The right histogram shows the distribution of the generators' dimensions.}
\label{fig:F8data}
\end{figure}

We generated a smaller dataset containing 1,584,099 generalised linear codes over~$\FF_8$; the dataset is summarised in Figure~\ref{fig:F8data}. The complexity for predicting codes over~$\FF_8$ is typically larger than over~$\FF_7$, and the amount of data we collected was marginally smaller than for~$\FF_7$. To mitigate this, we used a two-stage approach to training the model: first, to extract useful features of codes from the common examples in the pre-training stage; and second, to generalise this knowledge to all the classes in the training stage. Pre-training was done on a dataset with around 300,000 examples, consisting of classes with more than 10,000 examples. In the training stage, we used all the classes, down-sampling or over-sampling all available classes to 600, with a total dataset of 20,000 elements. We used 9:1 train/test split in both stages.

\subsection*{Model architecture}
We formulated the prediction of the minimum Hamming distance as an~NLP-style sequence classification task. The input sequence come from interpreting the canonical generator matrix~$G$ of a generalised toric code as a sequence~$S$ of length $k$ whose elements (the rows of~$G$) are codewords in $\FF_q^{(q-1)^2}$, and the class label is the minimum Hamming distance~$d\in \{0, \dots, (q-1)^2\}$. The maximum value $d=(q-1)^2$ occurs only when $k=1$; hence we restrict the classes to~$\{0,\dots,(q-1)^2-1\}$.

\begin{figure}
\centering
 \includegraphics[width=0.35\textwidth]{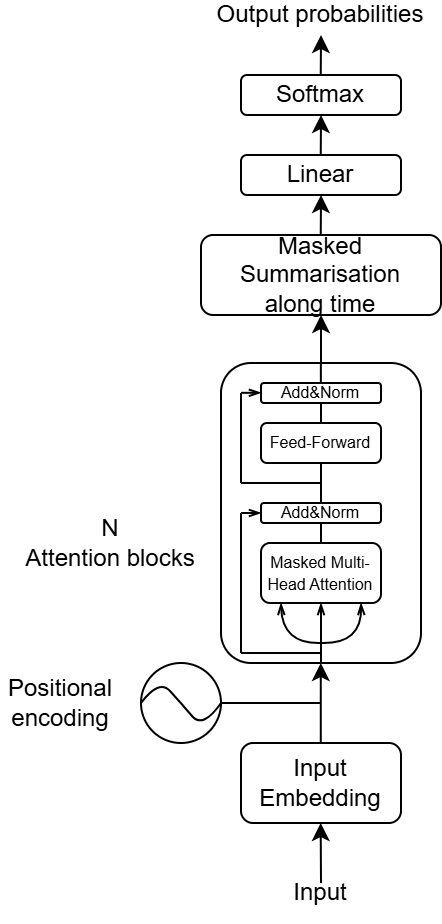}
 \caption{Toric transformer -- model architecture.}
 \label{fig:TransformerScheme}
\end{figure}

We modified a stripped-down GPT-2~\cite{radford2019language,karpathyGPT} to perform sequence classification. The adapted model uses: embeddings for field elements (none for~$\FF_7$, 4-dimensional vectors for~$\FF_8$); sinusoidal positional encodings of rows; two layers of attention blocks, each with masked self-attention and a feed-forward neural network; masked attention summarising information from each row; and a soft-max transformation to produce probabilities for minimum-distance classes. This is illustrated in Figure~\ref{fig:TransformerScheme}. See the Appendix for hyper-parameters used, the general structure of the network, and for a more detailed discussion.

\section{Results}\label{sec:results}
\subsection*{Training and performance}
To measure performance, we report the proportion of estimates falling within a three-unit range ($d\pm 3$) of the actual minimum Hamming distance~$d$ (we call this the~\emph{accuracy}), alongside the mean absolute error~(MAE) and mean squared error~(MSE) per minimum-distance class.

\begin{figure}[t]
\centering
\includegraphics[width=1\textwidth]{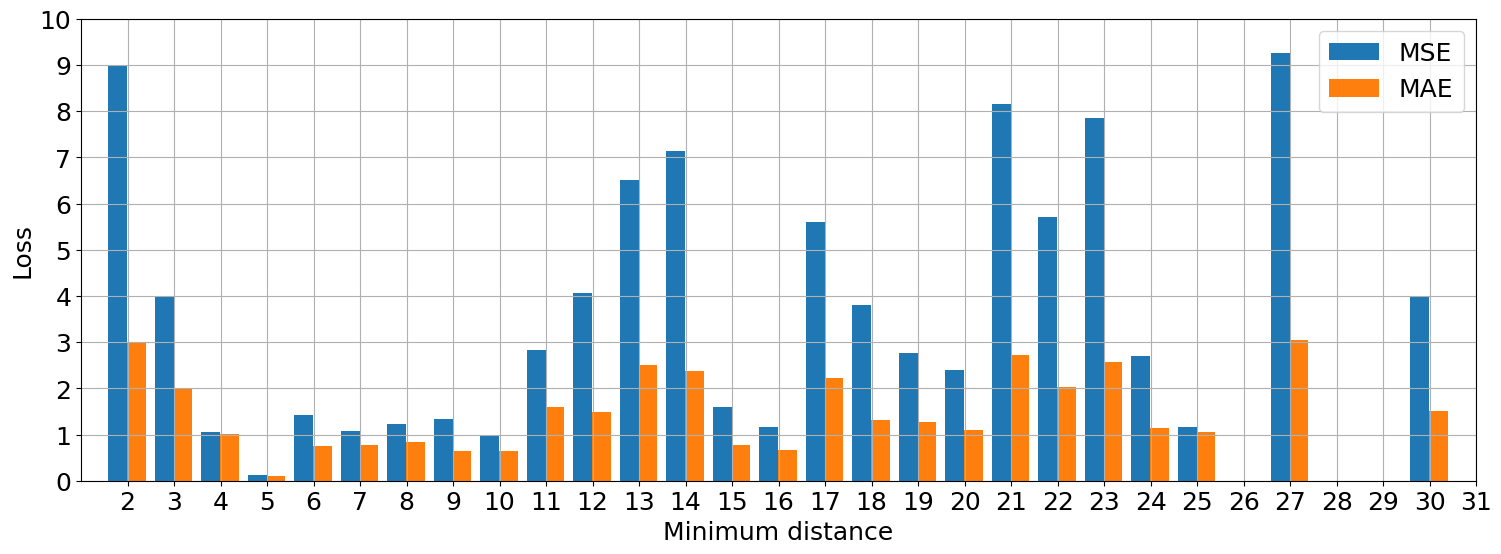}
 \caption{Losses on a test set for~$\FF_7$ codes.}
 \label{fig:F7performance}
\end{figure}

Over~$\FF_7$, our model's train set achieved overall~$1.04$ MAE and~$91.9\%$ accuracy. Per class~MAE and~MSE can be seen in Figure~\ref{fig:F7performance}. On the test set, consisting of around 61,000 examples, our model achieved~$1.05$ MAE and~$91.6\%$ accuracy. Notice that for all classes, the~MAE is less than three and most MSEs roughly equal the square of the~MAEs, indicating only a small number of larger errors are present in the model's predictions.

\begin{figure}[t]
\centering
\includegraphics[width=1\textwidth]{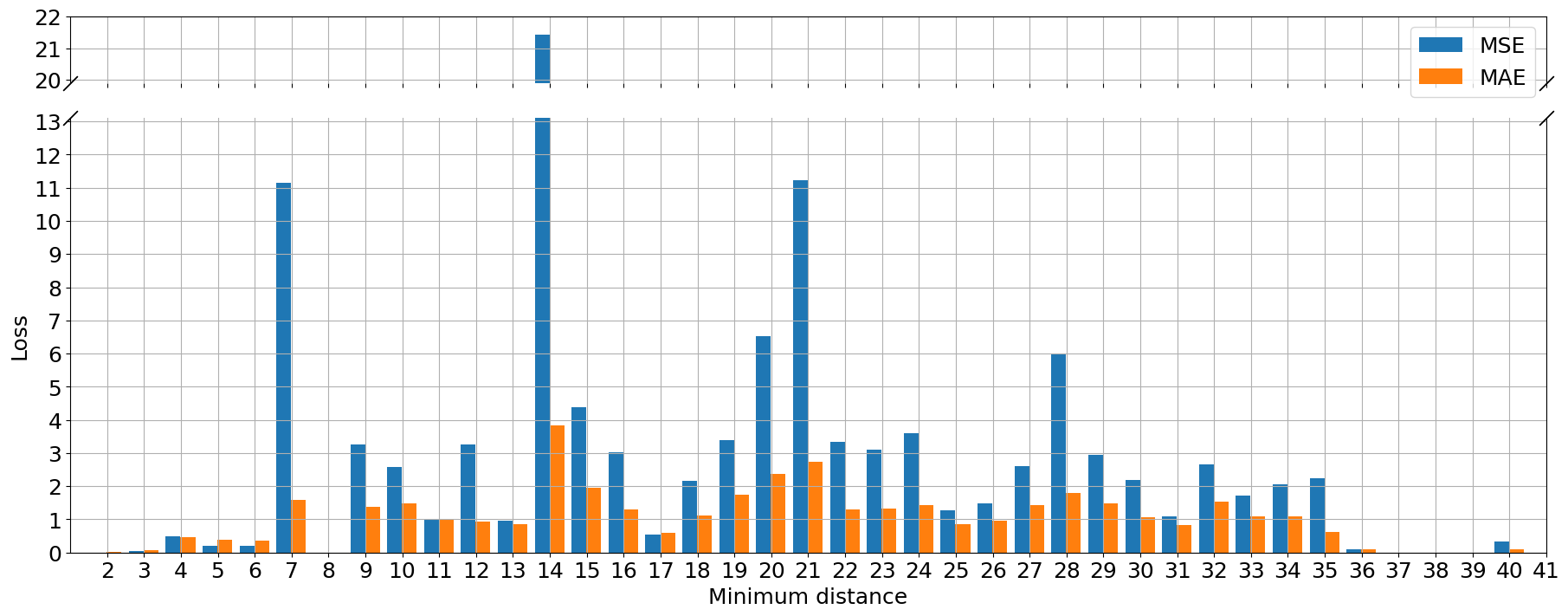}
 \caption{Losses on a test set for~$\FF_8$ codes.}
 \label{fig:F8performance}
\end{figure}

The~$\FF_8$ model's post-train stage achieved overall MAE~$1.09$ and~$93.4\%$ accuracy on the train set and MAE~$1.21$ and accuracy~$92.4\%$ on the test set. Figure~\ref{fig:F8performance} shows per class~MAE and~MSE measurements. Although these numbers trail the~$\FF_7$ results slightly, they remain competitive, and most minimum-distance classes stay well within the~$\pm 3$-error target. Larger deviations are confined to a few classes ($d=7,14,20,21,28$), showing opportunities for future improvement.

\subsection*{Evolving champion codes}
We now describe how we coupled our predictive model with a genetic algorithm~(GA) in order to search for optimal generalised toric codes by dimension. Repeated runs over~$\FF_7$ rediscovered champion codes seen in~\cite{kasprzyk1}; the same approach over~$\FF_8$ discovered new champion linear codes. This approach may be adapted to other classes of codes, including Reed–Muller codes~\cite{MullerD.E.1954AoBa, ReedI.1954Acom}, Bose–Chaudhuri–Hocquenghem codes~\cite{hocquenghem1959codes, BoseR.C.1960Oaco}, and algebrogeometric codes~\cite{couvreur:hal-02931167, VDGoppa1983}.

\subsection*{Genetic algorithms}\label{par:GA}
Fix a positive integer~$m$ and a field~$\FF_q$, and consider a \emph{population} $\{ V_i\}$ consisting of sets $V_i$ of $m$ lattice points in~$[0,q-2]^2$; in practise $m=k$ is our target dimension; --this is justified below. A set~$V=\{(a_1,b_1),\ldots,(a_m,b_m)\}$ is called a~\emph{chromosome}, and a lattice point~$(a_j, b_j)\in [0,q-2]^2$ is called a~\emph{gene}. With the fitness function of our genetic algorithm as our model's predicted minimum distance (see Algorithm~\ref{alg:fitnessfunction}), each generation, guided by a single parameter test with the models, proceeds as follows.
\begin{enumerate}
 \item \textbf{Selection.} Stochastic universal sampling picks 200 parents from a population of 300.
 \item \textbf{Crossover.} Flatten each parent’s chromosome into a list~$[a_1,b_1,\dots,a_m,b_m]$ of length~$2m$. A random crossover split point made between genes swaps the tails; offspring with duplicate lattice points~$(a_i,b_i)=(a_j,b_j)$, $i\neq j$, are discarded and we repeat with other parents.
 \item \textbf{Mutation.} With 10\% probability, a lattice point is replaced by a new, unused one, preserving~$m$ distinct lattice points.
 \item \textbf{Elitism.} The best 30 chromosomes carry forward unchanged, with the rest discarded.
\end{enumerate}
We repeat for~200 generations. The genetic algorithms we used can be obtained from~\cite{GAs}.

\begin{algorithm}[t]
\small
\DontPrintSemicolon
\SetKwData{GeneratorMatrix}{GeneratorMatrix}
\SetKwData{Model}{Model}
\KwData{Target dimension~$k$; a chromosome~$V\subset [0,q-2]^2$ to be evaluated; a set of sets~$S$ containing the chromosomes found by the~GA so far.}
\KwResult{Updated value for~$S$ and a fitness score $f$ for $V$.}
\BlankLine
\eIf{$V\in S$}{
$f\leftarrow 10$\;
}{
$k_C\leftarrow\dim C_V(\FF_q)$\;
$G\leftarrow\GeneratorMatrix(C_V(\FF_q))$
\tcp*{calculate the canonical generator matrix}
$d_\mathrm{approx}\leftarrow\Model(G)$
\tcp*{estimate the Hamming distance}
$S\leftarrow S\cup\{V\}$\;
$f\leftarrow 300+d_{\mathrm{approx}} - |k_C - k|$
}
\caption{Fitness function}\label{alg:fitnessfunction}
\end{algorithm}

Note that we evolve lattice point sets~$V$ rather than generator matrices~$G$: there is no clear way to mutate a generator matrix whilst preserving the class codes considered. By mutating~$V$ we can guarantee that we remain in the class of generalised toric codes. More generally, when there exists a way to mutate a code whilst remaining within a specified class, we say that that class has an~\emph{evolvable parameter space}. Spaces of this sort include: generalised toric codes; Reed--Muller codes; Bose--Chaudhuri--Hocquenghem codes; and algebrogeometric codes. It could be possible that certain classes of quantum codes also have an evolvable parameter space.

We do not store all codes generated during a run of the~GA; only those produced for~BZ evaluation will be stored in our datasets. We target champion codes of dimension~$k$ by setting~$m=k$ and penalising candidates whose dimension drifts from~$k$. To see why the dimension~$k$ should equal the number of lattice points~$m$, observe that: on the one hand, increasing~$m$ does not improve our chances of producing a champion code of dimension~$k$; on the other hand, we must have~$k \leq m$. Thus, optimal codes should be found when~$k = m$. Additional penalties prevent rediscovering codes found in earlier runs. This leads to the fitness function described in Algorithm~\ref{alg:fitnessfunction}, and used over both~$\FF_7$ and~$\FF_8$; note that the number 300 in the fitness score is simply a way to offset the other two terms and ensure that the output of the formula is positive, as required by the~\texttt{PyGAD} library~\cite{gad2023pygad}. During~GA search, we predict each code’s minimum Hamming distance with our model ($d_{\mathrm{approx}}$); promising candidates are then verified with the computationally expensive~BZ algorithm. Since evaluating the minimum Hamming distance is the main bottleneck, and this is much costlier to check over~$\FF_8$, we apply the~BZ check more selectively in that case.

\subsection*{Champion codes over~$\FF_7$}
We applied our~$\FF_7$ model to rediscover the best generalised toric codes over~$\FF_7$ from~\cite{kasprzyk1}; see Algorithm~\ref{alg:cap}.

\begin{algorithm}[t]
\small
\DontPrintSemicolon
\SetKwData{GA}{GA}
\SetKwData{BZ}{BZ}
\SetKwData{UpdateDistances}{UpdateDistances}
\SetKwProg{Def}{def}{:}{}
\KwData{A list~$(d_k)$ of the largest known minimum Hamming distance~$d_k$ in dimension~$k$.}
\KwResult{A list~$(d'_k)$ of the largest minimum Hamming distance~$d'_k$ discovered by the~GA in dimension~$k$; a list~$(S_k)$ of discovered codes in dimension~$k$ with minimum Hamming distance $d'_k$.}
\BlankLine
\Def{UpdateDistances($S$, $k$)}{
\For{$V\in S\setminus S^\mathrm{seen}$}{
$d_C\leftarrow\BZ(C_V(\FF_7))$
\tcp*{calculate the minimum Hamming distance}
\uIf{$d_C=d'_k$}{
$S_k\leftarrow S_k\cup\{V\}$\;
}
\ElseIf{$d_C>d'_k$}{
$d'_k,S_k\leftarrow d_C,\{V\}$\;
}
}
$S^\mathrm{seen}\leftarrow S^\mathrm{seen}\cup S$\;
}
\BlankLine
\emph{initialise the data}\;
$S^\mathrm{seen}\leftarrow\{\}$\;
\For{$k\leftarrow 3$ \KwTo $34$}{
$d'_k,S_k\leftarrow 0,\{\}$\;
}
\BlankLine
\emph{iteratively update the distances}\;
\Repeat{some termination condition is satisfied}{
\For{$k\leftarrow 3$ \KwTo $34$}{
\If{$d'_k\neq d_k$}{
$S\leftarrow\GA(k)$
\tcp*{generate codes of dimension~$k$}
\UpdateDistances{$S$, $k$}\;
}
}
}
\caption{Discovering best codes over~$\FF_7$ for dimensions~$3\leq k \leq 34$}\label{alg:cap}
\end{algorithm}

Across 757~GA runs (see Table~\ref{tab:dim_vs_runs}), most dimensions yielded a best-distance code on the first attempt. Dimensions~12 and~15 needed more iterations, indicating a tougher search landscape. Dimension~19 remains elusive after~501 runs, pointing to the rarity of~$[36,19,12]$ generalised toric codes likely coupled with a tougher search landscape. 

\begin{table}[t]
 \centering
 \small
 \begin{tabular}[t]{ccc}
 \toprule
$\bm{k}$ & \textbf{\#Runs} & \textbf{\#Codes}\\
 \midrule
 3 & 1 & 98\\
 4 & 1 & 332\\
 5 & 1 & 23446\\
 6 & 1 & 4023\\
 7 & 2 & 115\\
 8 & 1 & 564\\
 9 & 7 & 27\\
 10 & 1 & 2482\\
 11 & 3 & 124\\
 12 & 55 & 0\\
 13 & 1 & 411\\
 \bottomrule
 \end{tabular}
 \qquad
 \begin{tabular}[t]{ccc}
 \toprule
$\bm{k}$ & \textbf{\#Runs} & \textbf{\#Codes}\\
 \midrule
 14 & 2 & 16\\
 15 & 82 & 1\\
 16 & 1 & 11838\\
 17 & 1 & 1951\\
 18 & 20 & 66\\
 19 & n/a & n/a \\
 20 & 1 & 1094\\
 21 & 15 & 39\\
 22 & 18 & 57\\
 23 & 1 & 336\\
 24 & 43 & 5\\
 \bottomrule
 \end{tabular}
 \qquad
 \begin{tabular}[t]{ccc}
 \toprule
$\bm{k}$ & \textbf{\#Runs} & \textbf{\#Codes}\\
 \midrule
 25 & 1 & 45714\\
 26 & 1 & 21432\\
 27 & 1 & 4215\\
 28 & 6 & 80\\
 29 & 1 & 2420\\
 30 & 1 & 33302\\
 31 & 1 & 10713\\
 32 & 1 & 0\\
 33 & 1 & 0\\
 34 & 1 & 0\\
 \bottomrule
 \end{tabular}
 \caption{The dimension~$k$ of the code, the number of iterations required to find the best minimum Hamming distance, and the number of codes found that obtain the minimum Hamming distance, over~$\FF_7$.}
 \label{tab:dim_vs_runs} 
\end{table}

Table~\ref{tab:dim_vs_runs} shows a loose correlation: dimensions that contain a larger number of best codes in the dataset typically require fewer~GA runs to rediscover them. This pattern fades at the smallest and largest dimensions, where optimal codes are naturally easier to locate than those in the 12–28 range. Thus, dimensions 12, 15, 19, and 28 converged more slowly because the dataset held few optimal-code examples at those sizes (this is amplified for dimensions 19 and 28 by the rarity of examples in the training set). Notably, though there were no best codes of dimension 12 in the training and testing datasets, our method still found a best generalised toric code of dimension 12. We see this more apparently over~$\FF_8$, where we discover new champion codes.

\subsection*{Champion codes over~$\FF_8$}
Because champion codes over~$\FF_8$ are largely unknown, we screen candidates against the best-known lower bounds~$d_k$ from~\cite{codetables}. For a code~$C$ of dimension~$k$, we use~\texttt{Magma}’s \texttt{VerifyMinimumDistanceLowerBound}$(C,d_k)$ intrinsic. If the~BZ routine shows~$d_C<d_k$, we discard~$C$ immediately; otherwise we keep it for a full distance check. This filter saves considerable computation by focusing effort only on codes that~\emph{could} beat the current records. See Algorithm~\ref{alg:cap2}.

\begin{algorithm}[t]
\small
\DontPrintSemicolon
\SetKwData{GA}{GA}
\SetKwData{BZ}{BZ}
\SetKwData{VerifyMinimumDistanceLowerBound}{VerifyMinimumDistanceLowerBound}
\SetKwData{UpdateCandidates}{UpdateCandidates}
\SetKwProg{Def}{def}{:}{}
\KwData{A list~$(d_k)$ of the best known lower bound on the minimum Hamming distance~$d_k$ in dimension~$k$.}
\KwResult{A list $(d'_k)$ of the largest minimum Hamming distance $d'_k\geq d_k$ discovered by the~GA in dimension~$k$; a list~$(S_k)$ of discovered codes in dimension~$k$ with minimum Hamming distance $d'_k$.}
\BlankLine
\Def{UpdateCandidates($S$, $k$)}{
\For{$V\in S\setminus S^\mathrm{seen}$}{
$C\leftarrow C_V(\FF_8)$\;
\If{\VerifyMinimumDistanceLowerBound($C$, $d_k$)}{
$S^\mathrm{candidate}_k\leftarrow S_k\cup\{V\}$\;
}
}
$S^\mathrm{seen}\leftarrow S^\mathrm{seen}\cup S$\;
}
\BlankLine
\emph{initialise the data}\;
$S^\mathrm{seen}\leftarrow\{\}$\;
\For{$k\leftarrow 3$ \KwTo $46$}{
$S^\mathrm{candidate}_k\leftarrow \{\}$\;
}
\BlankLine
\emph{iteratively update the candidates}\;
\Repeat{some termination condiion is satisfied}{
\For{$k\leftarrow 3$ \KwTo $46$}{
$S\leftarrow\GA(k)$
\tcp*{generate codes of dimension~$k$}
\UpdateCandidates($S$, $k$)\;
}
}
\BlankLine
\emph{verify the candidates}\;
\For{$k\leftarrow 3$ \KwTo $46$}{
$d'_k, S_k\leftarrow d_k, \{\}$\;
\For{$V\in S^\mathrm{candidate}_k$}{
$d_C\leftarrow\BZ(C_V(\FF_8))$
\tcp*{calculate the minimum Hamming distance}
\uIf{$d_C=d'_k$}{
$S_k\leftarrow S_k\cup\{V\}$\;
}
\ElseIf{$d_C>d'_k$}{
$d'_k,S_k\leftarrow d_C,\{V\}$\;
}
}
}
\caption{Discovering new champion codes over~$\FF_8$ for dimensions~$3 \leq k \leq 46$}\label{alg:cap2}
\end{algorithm}

We performed one run, due to time and computational constraints, of the above algorithm, excluding dimension 30 due to the high computational cost of running the~BZ algorithm on this dimension. According to~\texttt{Magma}'s time estimates, computing the 300 codes for dimension 30 would have exceeded our computational time budget. Hence there is no~GA run for~$l=30$ in Table~\ref{tab:efficiencyF8}. Despite this, we found over 700 possibly new champion codes. As with~$\FF_7$, we compared these results with the number of champion codes found in the dataset on which the~$\FF_8$ model trained. Once again, champion codes were discovered at the lowest and highest dimensions for similar reasons as over~$\FF_7$. Rather surprisingly, in only one run, we achieved champion codes in dimensions that had no champion codes in their datasets: dimensions 18, 22, and 38. Through explicit calculation compared with the known best codes in~\cite{codetables}, we see that all codes of dimension 18, 22, and 38 that have been discovered are in fact new: see Table~\ref{tab:182238examplecodes}. 

Note that we had previously run the algorithm for $20\leq m\leq 29$, where $m$ is the size of the lattice point sets considered. In this search, we discovered a champion code of dimension 22, as seen in Table~\ref{tab:example}. Upon discovery, we saved this code, however due to an error on our part we were unable to save the corresponding dataset that included this code. As a result, this code is~\emph{not} noted in Table~\ref{tab:efficiencyF8}.

For any linear code, a standard upper bound on the minimum distance is given by the Singleton bound, $d \le n - k + 1$. A comparison with our results (for~$n = 49$) appears in Table~\ref{tab:knowncodebounds}. See~\cite{OrderBoundforToricCodes,little1,minkowskilength,little2006toric} for sharper bounds based on additional code information.

\begin{table}[t]
\centering
\small
\begin{tabular}{r@{}p{2mm}@{\ }cccccccccccc}
\toprule
$\bm{k}$&&3&4&6&7&8&18&22&38&40&41&44&46\\
\cmidrule{1-1}\cmidrule{3-14}
\textbf{Min.\ dist.}&&42&40&36&35&34&21&18&7&6&6&4&3\\
\textbf{Singleton bound}&&47&46&44&43&42&32&28&12&10&9&6&4\\
\bottomrule
\end{tabular}
\caption{Code parameters with~$n = 49$ and corresponding Singleton bounds.}
\label{tab:knowncodebounds}
\end{table}

\section{Comparing our method with random search}
Recall that our method aims to minimise the number of~BZ evaluations required to identify new champion codes. To illustrate our method’s advantage over random search, we first detail the size of the search space. Over~$\FF_q$, there are~$2^{(q-1)^2}$ possible lattice point sets for constructing generalised toric codes. However, distinct sets~$V_1\neq V_2$ may yield equivalent linear codes. We approximate the number of inequivalent codes as
\begin{equation*}
\left|\frac{X}{\mathrm{AGL}_2(\FF_q)}\right|
 \approx \frac{2^{(q-1)^2}}{q^2(q^2-1)(q^2-q)},
\end{equation*}
where~${\mathrm{AGL}_2(\FF_q)}$ is the affine general linear group of~$\FF_q$ in two dimensions. One may derive this formula as follows. By Burnside's Lemma,
\begin{equation*}
 \left|\frac{X}{\mathrm{AGL}_2(\FF_q)}\right| = \frac{1}{|\mathrm{AGL}_2(\FF_q)|}\sum_{g \in \mathrm{AGL}_2(\FF_q)}|X^g|, 
\end{equation*}
where~$X^g = \{ x\in X: g\cdot x = x\}$. Now,
\begin{align*}
|\mathrm{AGL}_2(\FF_q)| &= |\FF_q^2| \cdot |\mathrm{GL}_2(\FF_q)|\\
&= q^2 (q^2-1)(q^2-q).
\end{align*}
Consider~$|X^g|$ for each~$g \in \mathrm{AGL}_2(\FF_q)$: if~$g$ is the identity, then all configurations are fixed, so~$|X^g| = 2^{(q-1)^2}$; if~$g$ is not the identity, then we have significantly fewer fixed configurations, which we may approximate as negligible as~$q$ increases.

Approximate values for the number of unique generalised toric codes over~$\FF_q$ ($q = 7,8,9,11,13$) are shown in Table~\ref{tab:sizeofspace}. Note that this estimate is less accurate for small~$q$, where fixed configurations under affine transformations are non-negligible, but becomes effective for large~$q$, where exact enumeration is infeasible. To our knowledge, no prior work has quantified the proportion of generalised toric codes with a given minimum Hamming distance among codes of fixed dimension or $|V|$.

\begin{table}[t]
\centering
\small
\begin{tabular}{r@{}p{2mm}@{\ }ccccc}
\toprule
$\bm{q}$&&7&8&9&11&13\\
\cmidrule{1-1}\cmidrule{3-7}
\textbf{Approx.\ size}&&$6.96 \cdot 10^{5}$&$2.49 \cdot 10^{9}$&$3.95 \cdot 10^{13}$&$7.94 \cdot 10^{23}$&$5.03 \cdot 10^{36}$\\
\bottomrule
\end{tabular}
\caption{Approximate size of~$|X/\mathrm{AGL}_2(\FF_q)|$}
\label{tab:sizeofspace}
\end{table}

One way to compare the efficiency of the~GA over a random search is to estimate the expected number of~BZ evaluations required to find a champion code. Let~$N_k^{(m)}$ denote the number of distinct codes of dimension~$k$ generated from lattice point sets of size $m$, and let~$C_k^{(m)}$ denote the number of champions amongst them. The expected number of evaluations in a random search is given by
\begin{equation*}
 \mathbb{E} [ \text{steps before champion} ] = \frac{ C_k^{(m)} }{ N^{(m)} },
 \qquad
 \text{ where }N^{(m)} = \sum_{k} N_k^{(m)}.
\end{equation*}
For the~GA, this expectation can be estimated using the number of codes evaluated and champions found,~$\widetilde{C}_k^{(m)}$, giving
\begin{equation*}
 \mathbb{E} [ \text{steps before champion for~GA} ] = \frac{ \widetilde{C}_k^{(m)} }{ N^{(m)} }.
\end{equation*}
Tables~\ref{tab:efficiencyF7} and~\ref{tab:efficiencyF8} summarise these statistics for both random search and our~GA search over~$\FF_7$ and~$\FF_8$.

For smaller and larger values of~$m$, where champions are common, using the~GA provides little advantage. However, these cases are not computationally difficult. In contrast, for intermediate values of~$m$ the~GA achieves up to a twofold reduction in~BZ evaluations relative to random search. This can be attributed to two factors:
\begin{enumerate}
 \item in certain dimensions (such as~$k=16$ over~$\FF_7$ and~$k=40$ over~$\FF_8$), the high density of champions causes the~GA to saturate once it retains 30 elite codes per generation; and
 \item in sparsely represented dimensions (such as~$16\leq k\leq 39$ over~$\FF_8$), few or no champions appear in the training data, limiting the~ML model’s predictive accuracy, so that champion codes receive low approximations from the model.
\end{enumerate}

An ideal case over~$\FF_7$ is $k=14$, where the dataset contained few champion codes, making random discovery difficult. Nevertheless, the~GA located a champion within two iterations. For~$\FF_8$, the mid-range~$18\leq m\leq 39$ is especially sparse of codes. Given the exponentially larger search space compared to~$\FF_7$, one might expect the discovery of champions to require substantially more~BZ evaluations, making the discovery of champions in dimensions 18, 22, and 38 particularly striking. Despite limited representation in the dataset, the~ML model successfully generalised from other dimensions to poorly represented ones, allowing our method to evolve new champion codes.

\section{Discussion}
Our method provides a new tool for the construction of champion error-correcting codes. We note some ways in which this approach can be improved: larger and more balanced datasets would give predictive models richer coverage, especially at rare minimum Hamming distances; broader hyper-parameter searches (optimisers, mixed cross-entropy and Wasserstein losses, alternative row/field encodings) and systematic embedding design could further improve accuracy; and improvements to the~BZ algorithm~\cite{stefka2021} can accelerate the discovery of champion codes by reducing the computational cost of determining the minimum Hamming distance.

We wish to quantitatively characterise the distribution of generalised toric codes with respect to their minimum Hamming distances for fixed dimension~$k$ or fixed lattice point set size~$m$ -- a question that has not yet been explored. The~GA itself invites tuning: a wider sweep of population sizes, crossover strategies, and penalty weights, plus multi-week runs over~$\FF_8$ with staged~\texttt{Magma} checks, should surface additional champions. It may also be fruitful to consider other ways to investigate the parameter space~\cite{SipperMoshe2018Itps}.

The same framework will apply essentially as-is to larger prime-powered finite fields~$\FF_q$, $q>8$; it can be expanded to (non-generalised) toric codes, where we can exploit algebrogeometric properties arising from their toric description~\cite{jhansen1, little2006toric, minkowskilength}; and it can be adapted to construct possible champion codes from other families with an evolvable parameter space (BCH, Reed–Muller, and more), once suitable~GA `genes' are defined. A different approach due to Vaessens et al.~\cite{VAESSENSR.J.M1993Gaic} gives a general~GA scheme for building large codes, but this would be unsuitable for searching for codes of a fixed dimension as the fitness function would require a dimension penalty, as done in Algorithm~\ref{alg:fitnessfunction}. Finally, recasting the task as an image classification problem with padded matrices, or applying representation-learning techniques from NLP~\cite{bengio2013representation}, offer alternative modelling avenues worth exploring.

\begin{table}[t]
\centering
\small
\begin{tabular}[t]{cc}
\toprule
$\bm{k}$ & \textbf{\#Champions} \\
\midrule
1 & 0 \\
2 & 0 \\
3 & 16464 \\
4 & 31091 \\
5 & 0 \\
6 & 1034 \\
7 & 216 \\
8 & 28 \\
9 & 3 \\
10 & 11 \\
11 & 1 \\
12 & 144 \\
13 & 1 \\
14 & 0 \\
15 & 146 \\
\bottomrule
\end{tabular} 
\qquad
\begin{tabular}[t]{cc}
\toprule
$\bm{k}$ & \textbf{\#Champions} \\
\midrule
16 & 1 \\
17 & 0 \\
$\vdots$ & $\vdots$ \\
39 & 0 \\
40 & 6138 \\
41 & 357 \\
42 & 0 \\
43 & 0 \\
44 & 5177 \\
45 & 0 \\
46 & 16464 \\
47 & 0 \\
48 & 1176 \\
\bottomrule
\end{tabular} 
\caption{The dimension~$k$ and number of champion codes in training and testing datasets, over~$\FF_8$.}
\label{tab:dim_vs_bestF82} 
\end{table}

\begin{table}[t]
 \centering
 \small
 \begin{tabularx}{\textwidth}{cc@{}p{3mm}@{}c@{}p{3mm}@{}X}
 \toprule
$\bm{k}$ & $\bm{d}$ && $\bm{N}$ && \textbf{Lattice points}~$\bm{V}$ \\
\cmidrule{1-2}\cmidrule{4-4}\cmidrule{6-6}
3 & 42 && 231 && \mbox{(1, 0)}, \mbox{(2, 3)}, \mbox{(6, 3)} \\
4 & 40 && 66 && \mbox{(0, 5)}, \mbox{(2, 6)}, \mbox{(6, 0)}, \mbox{(6, 4)} \\
6 & 36 && 3 && \mbox{(0, 0)}, \mbox{(0, 6)}, \mbox{(2, 4)}, \mbox{(2, 5)}, \mbox{(4, 6)}, \mbox{(6, 4)} \\
7 & 35 && 1 && \mbox{(0, 3)}, \mbox{(2, 5)}, \mbox{(4, 6)}, \mbox{(5, 0)}, \mbox{(5, 1)}, \mbox{(6, 0)}, \mbox{(6, 3)} \\
8 & 34 && 1 && \mbox{(0, 4)}, \mbox{(0, 6)}, \mbox{(2, 3)}, \mbox{(3, 4)}, \mbox{(4, 0)}, \mbox{(5, 1)}, \mbox{(5, 3)}, \mbox{(6, 0)} \\
18 & 21 && 4 && \mbox{(0, 1)}, \mbox{(0, 2)}, \mbox{(0, 3)}, \mbox{(0, 5)}, \mbox{(1, 0)}, \mbox{(1, 1)}, \mbox{(1, 5)}, \mbox{(1, 6)}, \mbox{(3, 2)}, \mbox{(3, 3)}, \mbox{(3, 5)}, \mbox{(3, 6)}, \mbox{(4, 3)}, \mbox{(4, 4)}, \mbox{(5, 1)}, \mbox{(5, 3)}, \mbox{(5, 5)}, \mbox{(6, 4)} \\
22 & 18 && 1 && \mbox{(0, 0)}, \mbox{(0, 6)}, \mbox{(1, 0)}, \mbox{(1, 3)}, \mbox{(1, 4)}, \mbox{(1, 5)}, \mbox{(2, 0)}, \mbox{(2, 5)}, \mbox{(2, 6)}, \mbox{(3, 2)}, \mbox{(3, 3)}, \mbox{(3, 5)}, \mbox{(3, 6)}, \mbox{(4, 2)}, \mbox{(5, 0)}, \mbox{(5, 3)}, \mbox{(5, 4)}, \mbox{(6, 0)}, \mbox{(6, 1)}, \mbox{(6, 3)}, \mbox{(6, 4)}, \mbox{(6, 5)} \\
38 & 7 && 1 && \mbox{(0, 1)}, \mbox{(0, 3)}, \mbox{(0, 4)}, \mbox{(0, 6)}, \mbox{(1, 1)}, \mbox{(1, 2)}, \mbox{(1, 3)}, \mbox{(1, 4)}, \mbox{(1, 5)}, \mbox{(1, 6)}, \mbox{(2, 0)}, \mbox{(2, 1)}, \mbox{(2, 3)}, \mbox{(2, 4)}, \mbox{(2, 6)}, \mbox{(3, 0)}, \mbox{(3, 1)}, \mbox{(3, 3)}, \mbox{(3, 4)}, \mbox{(3, 5)}, \mbox{(3, 6)}, \mbox{(4, 0)}, \mbox{(4, 1)}, \mbox{(4, 2)}, \mbox{(4, 3)}, \mbox{(4, 4)}, \mbox{(4, 5)}, \mbox{(4, 6)}, \mbox{(5, 0)}, \mbox{(5, 1)}, \mbox{(5, 3)}, \mbox{(5, 4)}, \mbox{(5, 5)}, \mbox{(5, 6)}, \mbox{(6, 0)}, \mbox{(6, 1)}, \mbox{(6, 2)}, \mbox{(6, 5)} \\
40 & 6 && 94 && \mbox{(0, 0)}, \mbox{(0, 2)}, \mbox{(0, 3)}, \mbox{(0, 4)}, \mbox{(0, 5)}, \mbox{(0, 6)}, \mbox{(1, 1)}, \mbox{(1, 2)}, \mbox{(1, 5)}, \mbox{(1, 6)}, \mbox{(2, 0)}, \mbox{(2, 1)}, \mbox{(2, 2)}, \mbox{(2, 3)}, \mbox{(2, 4)}, \mbox{(2, 5)}, \mbox{(2, 6)}, \mbox{(3, 0)}, \mbox{(3, 1)}, \mbox{(3, 2)}, \mbox{(3, 3)}, \mbox{(3, 4)}, \mbox{(3, 5)}, \mbox{(3, 6)}, \mbox{(4, 1)}, \mbox{(4, 2)}, \mbox{(4, 4)}, \mbox{(4, 5)}, \mbox{(4, 6)}, \mbox{(5, 1)}, \mbox{(5, 2)}, \mbox{(5, 3)}, \mbox{(5, 4)}, \mbox{(5, 5)}, \mbox{(5, 6)}, \mbox{(6, 0)}, \mbox{(6, 1)}, \mbox{(6, 4)}, \mbox{(6, 5)}, \mbox{(6, 6)} \\
41 & 6 && 3 && \mbox{(0, 0)}, \mbox{(0, 1)}, \mbox{(0, 2)}, \mbox{(0, 4)}, \mbox{(0, 5)}, \mbox{(0, 6)}, \mbox{(1, 0)}, \mbox{(1, 2)}, \mbox{(1, 3)}, \mbox{(1, 4)}, \mbox{(1, 5)}, \mbox{(1, 6)}, \mbox{(2, 0)}, \mbox{(2, 1)}, \mbox{(2, 2)}, \mbox{(2, 4)}, \mbox{(2, 6)}, \mbox{(3, 0)}, \mbox{(3, 1)}, \mbox{(3, 2)}, \mbox{(3, 3)}, \mbox{(3, 4)}, \mbox{(3, 5)}, \mbox{(3, 6)}, \mbox{(4, 0)}, \mbox{(4, 1)}, \mbox{(4, 4)}, \mbox{(4, 6)}, \mbox{(5, 0)}, \mbox{(5, 1)}, \mbox{(5, 2)}, \mbox{(5, 3)}, \mbox{(5, 5)}, \mbox{(5, 6)}, \mbox{(6, 0)}, \mbox{(6, 1)}, \mbox{(6, 2)}, \mbox{(6, 3)}, \mbox{(6, 4)}, \mbox{(6, 5)}, \mbox{(6, 6)} \\
44 & 4 && 100 && \mbox{(0, 0)}, \mbox{(0, 1)}, \mbox{(0, 2)}, \mbox{(0, 3)}, \mbox{(0, 4)}, \mbox{(0, 5)}, \mbox{(0, 6)}, \mbox{(1, 0)}, \mbox{(1, 1)}, \mbox{(1, 3)}, \mbox{(1, 4)}, \mbox{(1, 5)}, \mbox{(1, 6)}, \mbox{(2, 0)}, \mbox{(2, 1)}, \mbox{(2, 2)}, \mbox{(2, 3)}, \mbox{(2, 4)}, \mbox{(2, 5)}, \mbox{(2, 6)}, \mbox{(3, 0)}, \mbox{(3, 1)}, \mbox{(3, 4)}, \mbox{(3, 5)}, \mbox{(3, 6)}, \mbox{(4, 0)}, \mbox{(4, 1)}, \mbox{(4, 2)}, \mbox{(4, 3)}, \mbox{(4, 4)}, \mbox{(4, 6)}, \mbox{(5, 0)}, \mbox{(5, 1)}, \mbox{(5, 2)}, \mbox{(5, 3)}, \mbox{(5, 4)}, \mbox{(5, 5)}, \mbox{(6, 0)}, \mbox{(6, 1)}, \mbox{(6, 2)}, \mbox{(6, 3)}, \mbox{(6, 4)}, \mbox{(6, 5)}, \mbox{(6, 6)} \\
46 & 3 && 235 && \mbox{(0, 0)}, \mbox{(0, 1)}, \mbox{(0, 2)}, \mbox{(0, 3)}, \mbox{(0, 4)}, \mbox{(0, 5)}, \mbox{(0, 6)}, \mbox{(1, 0)}, \mbox{(1, 1)}, \mbox{(1, 2)}, \mbox{(1, 3)}, \mbox{(1, 4)}, \mbox{(1, 5)}, \mbox{(1, 6)}, \mbox{(2, 1)}, \mbox{(2, 2)}, \mbox{(2, 3)}, \mbox{(2, 4)}, \mbox{(2, 5)}, \mbox{(2, 6)}, \mbox{(3, 0)}, \mbox{(3, 1)}, \mbox{(3, 2)}, \mbox{(3, 3)}, \mbox{(3, 4)}, \mbox{(3, 6)}, \mbox{(4, 0)}, \mbox{(4, 1)}, \mbox{(4, 2)}, \mbox{(4, 3)}, \mbox{(4, 4)}, \mbox{(4, 5)}, \mbox{(4, 6)}, \mbox{(5, 1)}, \mbox{(5, 2)}, \mbox{(5, 3)}, \mbox{(5, 4)}, \mbox{(5, 5)}, \mbox{(5, 6)}, \mbox{(6, 0)}, \mbox{(6, 1)}, \mbox{(6, 2)}, \mbox{(6, 3)}, \mbox{(6, 4)}, \mbox{(6, 5)}, \mbox{(6, 6)}\\
\bottomrule
\end{tabularx}
\caption{Champion generalised toric codes found over~$\FF_8$ with block length~$n = 49$, dimension~$k$, and minimum Hamming distance~$d$. In each case, the number of such champion codes discovered is recorded by~$N$, and a representative set of lattice points is given by~$V$.} \label{tab:example}
\end{table}

\begin{table}[t]
 \centering
 \small
 \begin{tabularx}{\textwidth}{cc@{}p{3mm}@{}X}
 \toprule
$\bm{k}$ & $\bm{d}$ && \textbf{Lattice points}~$\bm{V}$ \\
\cmidrule{1-2}\cmidrule{4-4}
18 & 21 && \mbox{(0, 1)}, \mbox{(0, 2)}, \mbox{(0, 3)}, \mbox{(0, 5)}, \mbox{(1, 0)}, \mbox{(1, 1)}, \mbox{(1, 5)}, \mbox{(1, 6)}, \mbox{(3, 2)}, \mbox{(3, 3)}, \mbox{(3, 5)}, \mbox{(3, 6)}, \mbox{(4, 3)}, \mbox{(4, 4)}, \mbox{(5, 1)}, \mbox{(5, 3)}, \mbox{(5, 5)}, \mbox{(6, 4)} \\
18 & 21 && \mbox{(0, 1)}, \mbox{(0, 5)}, \mbox{(1, 3)}, \mbox{(1, 4)}, \mbox{(1, 6)}, \mbox{(2, 1)}, \mbox{(2, 5)}, \mbox{(2, 6)}, \mbox{(2, 7)}, \mbox{(3, 0)}, \mbox{(3, 1)}, \mbox{(3, 4)}, \mbox{(3, 6)}, \mbox{(4, 5)}, \mbox{(5, 1)}, \mbox{(6, 0)}, \mbox{(6, 1)}, \mbox{(6, 5)} \\
18 & 21 && \mbox{(0, 0)}, \mbox{(0, 1)}, \mbox{(0, 4)}, \mbox{(1, 1)}, \mbox{(1, 2)}, \mbox{(2, 6)}, \mbox{(3, 4)}, \mbox{(3, 6)}, \mbox{(4, 1)}, \mbox{(4, 2)}, \mbox{(4, 5)}, \mbox{(4, 6)}, \mbox{(5, 2)}, \mbox{(5, 5)}, \mbox{(6, 0)}, \mbox{(6, 4)}, \mbox{(6, 5)}, \mbox{(7, 6)} \\
18 & 21 && \mbox{(0, 1)}, \mbox{(0, 2)}, \mbox{(1, 3)}, \mbox{(1, 4)}, \mbox{(1, 6)}, \mbox{(2, 4)}, \mbox{(3, 2)}, \mbox{(3, 3)}, \mbox{(3, 4)}, \mbox{(3, 5)}, \mbox{(3, 6)}, \mbox{(4, 1)}, \mbox{(4, 2)}, \mbox{(4, 6)}, \mbox{(5, 1)}, \mbox{(5, 3)}, \mbox{(5, 5)}, \mbox{(6, 4)} \\
22 & 18 && \mbox{(0, 0)}, \mbox{(0, 6)}, \mbox{(1, 0)}, \mbox{(1, 3)}, \mbox{(1, 4)}, \mbox{(1, 5)}, \mbox{(2, 0)}, \mbox{(2, 5)}, \mbox{(2, 6)}, \mbox{(3, 2)}, \mbox{(3, 3)}, \mbox{(3, 5)}, \mbox{(3, 6)}, \mbox{(4, 2)}, \mbox{(5, 0)}, \mbox{(5, 3)}, \mbox{(5, 4)}, \mbox{(6, 0)}, \mbox{(6, 1)}, \mbox{(6, 3)}, \mbox{(6, 4)}, \mbox{(6, 5)} \\
38 & 7 && \mbox{(0, 1)}, \mbox{(0, 3)}, \mbox{(0, 4)}, \mbox{(0, 6)}, \mbox{(1, 1)}, \mbox{(1, 2)}, \mbox{(1, 3)}, \mbox{(1, 4)}, \mbox{(1, 5)}, \mbox{(1, 6)}, \mbox{(2, 0)}, \mbox{(2, 1)}, \mbox{(2, 3)}, \mbox{(2, 4)}, \mbox{(2, 6)}, \mbox{(3, 0)}, \mbox{(3, 1)}, \mbox{(3, 3)}, \mbox{(3, 4)}, \mbox{(3, 5)}, \mbox{(3, 6)}, \mbox{(4, 0)}, \mbox{(4, 1)}, \mbox{(4, 2)}, \mbox{(4, 3)}, \mbox{(4, 4)}, \mbox{(4, 5)}, \mbox{(4, 6)}, \mbox{(5, 0)}, \mbox{(5, 1)}, \mbox{(5, 3)}, \mbox{(5, 4)}, \mbox{(5, 5)}, \mbox{(5, 6)}, \mbox{(6, 0)}, \mbox{(6, 1)}, \mbox{(6, 2)}, \mbox{(6, 5)}\\
\bottomrule
\end{tabularx}
\caption{All champion generalised toric codes found over~$\FF_8$ with block length~$n=49$, dimension $k = 18, 22$, or~$38$, and minimum Hamming distance~$d$. In each case, the corresponding set of lattice points is given by~$V$.} \label{tab:182238examplecodes}
\end{table}

\begin{table}[t]
 \centering
 \small
 \begin{tabular}{c@{}p{3mm}@{}cccc@{}p{3mm}@{}cccc}
 \toprule
 &&
 \multicolumn{4}{c}{\textbf{Random}}&&
 \multicolumn{4}{c}{\textbf{GA}}\\
 &&&\textbf{\#Codes}&&\textbf{\#BZ per}&&&&&\textbf{\#BZ per}\\
 $\bm{m}$ &&
 \textbf{\#Codes} & \textbf{with $\bm{k=m}$} & \textbf{\#Champions} & \textbf{champion} &&
 \textbf{\#Runs} & \textbf{\#Codes} & \textbf{\#Champions} & \textbf{champion} \\
 \cmidrule{1-1}\cmidrule{3-6}\cmidrule{8-11}
 5 && 107086 & 86867 & 19307 & 4.50 && 1 & 300 & 72 & 4.17 \\
 6 && 108697 & 81094 & 3126 & 25.94 && 1 & 300 & 21 & 14.29 \\
 7 && 109942 & 74451 & 82 & 907.94 && 2 & 600 & 1 & 600.0 \\
 8 && 111957 & 67578 & 368 & 183.64 && 1 & 300 & 2 & 150.0 \\
 9 && 113803 & 60257 & 20 & 3012.85 && 7 & 2100 & 1 & 2100.0 \\
 10 && 115651 & 52704 & 1210 & 43.56 && 1 & 300 & 14 & 21.43 \\
 11 && 117155 & 45315 & 60 & 755.25 && 3 & 900 & 1 & 900.0 \\
 12 && 119072 & 38301 & 0 & n/a && 55 & 16500 & 1 & 16500.0 \\
 13 && 120182 & 31447 & 125 & 251.58 && 1 & 300 & 1 & 300.0 \\
 14 && 121730 & 25748 & 2 & 12874.0 && 2 & 600 & 1 & 600.0 \\
 15 && 120501 & 20857 & 1 & 20857.0 && 82 & 24600 & 1 & 24600.0 \\
 16 && 106083 & 16310 & 1859 & 8.77 && 1 & 300 & 32 & 9.38 \\
 17 && 74066 & 12458 & 350 & 35.59 && 1 & 300 & 7 & 42.86 \\
 18 && 32961 & 9026 & 22 & 410.27 && 4 & 1200 & 1 & 1200.0 \\
 19 && 6730 & 6730 & 0 & n/a && 84 & 25200 & 0 & n/a \\
 20 && 100000 & 100000 & 1094 & 91.41 && 1 & 300 & 4 & 75.0 \\
 21 && 100000 & 100000 & 39 & 2564.10 && 15 & 4500 & 1 & 4500.0 \\
 22 && 100000 & 100000 & 57 & 1754.39 && 18 & 5400 & 1 & 5400.0 \\
 23 && 100000 & 100000 & 336 & 297.62 && 1 & 300 & 1 & 300.0 \\
 24 && 100000 & 100000 & 5 & 20000.0 && 43 & 12900 & 1 & 12900.0 \\
 25 && 100000 & 100000 & 45714 & 2.19 && 1 & 300 & 134 & 2.24 \\
 26 && 100000 & 100000 & 21432 & 4.67 && 1 & 300 & 59 & 5.08 \\
 27 && 100000 & 100000 & 4215 & 23.72 && 1 & 300 & 10 & 30.0 \\
 28 && 100000 & 100000 & 80 & 1250.0 && 6 & 1800 & 1 & 1800.0 \\
 29 && 100000 & 100000 & 2420 & 41.32 && 1 & 300 & 6 & 50.0 \\
 30 && 100000 & 100000 & 33302 & 3.00 && 1 & 300 & 110 & 2.73 \\
 31 && 100000 & 100000 & 10713 & 9.33 && 1 & 300 & 33 & 9.09 \\
 \bottomrule
 \end{tabular}
 \caption{Search efficiency over~$\FF_7$ for generalised toric codes, comparing a random search with a~GA search. For each lattice point set size~$m$, the table records the total number of generated codes, those with dimension~$k = m$ (this is only relevant for the randomly generated codes), the number of champion codes discovered, and the average number of~BZ evaluations per champion.}.
 \label{tab:efficiencyF7}
\end{table}

\begin{table}[t]
 \centering
 \small
 \begin{tabular}{c@{}p{3mm}@{}cccc@{}p{3mm}@{}cccc}
 \toprule
 &&\multicolumn{4}{c}{\textbf{Random}}&&\multicolumn{4}{c}{\textbf{GA}}\\
 &&&&\textbf{\#BZ per}&\textbf{Min/max}&&&&\textbf{\#BZ per}&\textbf{Min/max}\\
 $\bm{k}$ &&
 \textbf{\#Codes} & \textbf{\#Champions} & \textbf{champion} & \textbf{distance} &&
 \textbf{\#Codes} & \textbf{\#Champions} & \textbf{champion} & \textbf{distance} \\
 \cmidrule{1-1}\cmidrule{3-6}\cmidrule{8-11}
 5 && 100000 & 0 & n/a & 36/38 && 300 & 0 & n/a & 35/38 \\
 6 && 100000 & 1034 & 96.71 & 36/36 && 300 & 3 & 100.0 & 36/36 \\
 7 && 100000 & 216 & 462.96 & 35/35 && 300 & 1 & 300.0 & 35/35 \\
 8 && 100000 & 28 & 3571.43 & 34/34 && 300 & 1 & 300.0 & 34/34 \\
 9 && 100000 & 3 & 33333.33 & 31/31 && 300 & 0 & n/a & 16/31 \\
 10 && 100000 & 11 & 9090.91 & 30/30 && 300 & 0 & n/a & 15/30 \\
 11 && 100000 & 1 & 100000.00 & 29/29 && 300 & 0 & n/a & 13/29 \\
 12 && 100000 & 144 & 694.44 & 28/28 && 300 & 0 & n/a & 12/28 \\
 13 && 100000 & 1 & 100000.00 & 27/27 && 300 & 0 & n/a & 11/27 \\
 14 && 100000 & 0 & n/a & 24/26 && 300 & 0 & n/a & 12/26 \\
 15 && 100000 & 146 & 684.93 & 24/24 && 300 & 0 & n/a & 16/24 \\
 16 && 101000 & 1 & 101000.0 & 24/24 && 300 & 0 & n/a & 10/24 \\
 17 && 100870 & 0 & n/a & 22/23 && 300 & 0 & n/a & 8/23 \\
 18 && 464 & 0 & n/a & 17/21 && 300 & 4 & 75.0 & 21/21 \\
 19 && 517 & 0 & n/a & 15/21 && 300 & 0 & n/a & 8/21 \\
 20 && 656 & 0 & n/a & 15/20 && 300 & 0 & n/a & 9/20 \\
 21 && 783 & 0 & n/a & 15/19 && 300 & 0 & n/a & 8/19 \\
 22 && 881 & 0 & n/a & 14/18 && 300 & 0 & n/a & 11/18 \\
 23 && 697 & 0 & n/a & 14/17 && 300 & 0 & n/a & 9/17 \\
 24 && 585 & 0 & n/a & 13/16 && 300 & 0 & n/a & 10/16 \\
 25 && 713 & 0 & n/a & 12/16 && 300 & 0 & n/a & 7/16 \\
 26 && 391 & 0 & n/a & 10/15 && 300 & 0 & n/a & 5/15 \\
 27 && 482 & 0 & n/a & 8/14 && 300 & 0 & n/a & 4/14 \\
 28 && 583 & 0 & n/a & 7/14 && 300 & 0 & n/a & 4/14 \\
 29 && 705 & 0 & n/a & 7/13 && 300 & 0 & n/a & 4/13 \\
 30 && 819 & 0 & n/a & 7/12 && n/a & n/a & n/a & 0/12 \\
 31 && 901 & 0 & n/a & 7/12 && 300 & 0 & n/a & 4/12 \\
 32 && 949 & 0 & n/a & 7/11 && 300 & 0 & n/a & 4/11 \\
 33 && 980 & 0 & n/a & 7/11 && 300 & 0 & n/a & 3/11 \\
 34 && 996 & 0 & n/a & 7/10 && 300 & 0 & n/a & 4/10 \\
 35 && 1000 & 0 & n/a & 7/9 && 300 & 0 & n/a & 4/9 \\
 36 && 1000 & 0 & n/a & 7/8 && 300 & 0 & n/a & 5/8 \\
 37 && 932 & 0 & n/a & 6/8 && 300 & 0 & n/a & 4/8 \\
 38 && 995 & 0 & n/a & 6/7 && 300 & 1 & 300.0 & 7/7 \\
 39 && 1000 & 0 & n/a & 6/7 && 300 & 0 & n/a & 3/7 \\
 40 && 21000 & 6138 & 3.42 & 6/6 && 300 & 95 & 3.16 & 6/6 \\
 41 && 41000 & 357 & 114.85 & 6/6 && 300 & 3 & 100.0 & 6/6 \\
 42 && 1000 & 0 & n/a & 5/6 && 300 & 0 & n/a & 3/6 \\
 43 && 21000 & 0 & n/a & 4/5 && 300 & 0 & n/a & 2/5 \\
 44 && 21000 & 5177 & 4.06 & 4/4 && 300 & 101 & 2.97 & 4/4 \\
 45 && 21000 & 0 & n/a & 3/4 && 300 & 0 & n/a & 2/4 \\
 46 && 18424 & 16464 & 1.12 & 3/3 && 300 & 236 & 1.27 & 3/3 \\
 \bottomrule
 \end{tabular}
 \caption{Search efficiency over~$\FF_8$ for generalised toric codes, comparing a random search with a~GA search. For each dimension~$k$, the table records the total number of generated codes, the number of champion codes, and the average~BZ evaluations per champion discovered. Note that we excluded~$k = 30$ from the~GA run due to the high computational cost of~BZ evaluations. Note also that a champion code of dimension 22, found by the~GA and given in Table~\ref{tab:example}, is omitted here, as noted in the section `Champion codes over~$\FF_8$'.}
 \label{tab:efficiencyF8}
\end{table}

\appendix
\section{Generalised toric codes}\label{app:sec1}
A~\emph{linear (error-correcting) code} over the finite field~$\FF_q$ is a~$k$-dimensional linear subspace~$C$ of the vector space~$\FF_q^n$; here~$q$ is a prime power. The vectors in~$C$ are called~\emph{codewords} of~$C$; the dimension~$n$ of the ambient vector space~$\FF_q^n$ is called the \emph{block length} of~$C$; and the number of possible codewords, $|C|=q^k$, is called the~\emph{size} of~$C$. Given a codeword $w=(w_1,\ldots,w_n)\in C$, the~\emph{weight} of $w$ is the number of non-zero entries $w_i$, that is, $|\{i : 1\leq i\leq n, w_i\neq 0\}|$. The~\emph{minimum Hamming distance} between two codewords $w,w'\in C$ is equal to the weight of $w-w'$, that is, $|\{i : 1\leq i\leq n, w_i\neq w'_i\}|$. The~\emph{minimum Hamming distance}~$d$ of the linear code is the minimum weight of its nonzero codewords or, equivalently, the minimum Hamming distance between any two distinct codewords. The block length~$n$, dimension~$k$, and minimum Hamming distance~$d$ of~$C$ are denoted by the triple~$[n,k,d]_q$, or simply by $[n,k,d]$ when the value of $q$ is clear. For an introduction to coding theory, see~\cite{abookoncodingtheory}.

Let $C$ be a linear code with parameters $[n,k,d]$. A choice of basis of $C$ gives rise to a $k\times n$~\emph{generator matrix}~$G$, which can be expressed in canonical form~$G = [I_k |H]$, where~$I_k$ is the~$k \times k$ identity matrix and~$H$ is a~$k \times (n-k)$ matrix. The dual generator matrix~$G^\perp$ of~$G$ is any matrix whose rows form a basis for the \emph{dual code}~$C^\perp$, where
\begin{equation*}
 C^\perp = \{v \in \FF_q^n : v \cdot w = 0 \text{ for all }w \in C\}.
\end{equation*}
The \emph{parity check} matrix~$P$ of~$G$ is an~$(n-k) \times n$ matrix satisfying~$GP^T = 0$. Since the rows of~$P$ generate~$C^\perp$,~$P$ is typically considered a dual generator matrix. Given the canonical form of the generator matrix, the canonical form of the parity check matrix~$P$ is~$[-H^T|I_{n-k}]$. Thus there is no ambiguity when considering the equivalent generator and parity check matrices of a linear code~$C$, and is standard to assume the canonical form when discussing the generator and parity check matrices.

The minimum Hamming distance~$d$ of a code~$C$ is a measure of the accuracy of the code for error correction. An important task is to find codes with~$d$ as large as possible for a given block length~$n$ and dimension~$k$. Theoretical upper and lower bounds on~$d$ are known, and Grassl~\cite{codetables} maintains a database of the current best known $d$, alongside a description of the corresponding code. We refer to any code with a larger minimum Hamming distance than the current best known example as a~\emph{champion} code, following the terminology of~\cite{kasprzyk1}. 

To calculate~$d$, we use the Brouwer--Zimmermann~(BZ) algorithm~\cite{zimmermann1996integral,grassl2006searching}. This algorithm generates several equivalent generator matrices, each with pivots located in a unique set of columns (known as the \emph{information set}). The relative rank of a generator matrix is defined as the number of pivot positions in its information set that are independent of previously constructed matrices. The algorithm involves enumerating all possible combinations of~$r$ generators for increasing values of~$r$. For each~$r$, if~$r$ becomes larger than the difference between the matrix's actual rank (i.e.\ its dimension) and its relative rank, the algorithm increases the lower bound on the minimum weight by one. Meanwhile, the upper bound on the minimum weight is determined by finding the smallest weight amongst the enumerated codewords. The computation completes once the lower and upper bounds converge. Despite several significant improvement~\cite{grassl2006searching}, the~BZ algorithm remains computationally costly. It has polynomial in~$k$ and exponential in~$q$ complexity~\cite{hernando2019algorithm}.

Hansen~\cite{hansen2000toric,jhansen1} introduced a special class of linear codes inspired by toric geometry. Given a lattice point $u=(a,b)\in [0,q-2]^2$, we associate a vector~$w_u\in\FF_q^{(q-1)^2}$ whose entries are given by evaluating the monomial~$x^ay^b$ at each point of the torus~$(\FF_q^*)^2$. More precisely, fix a primitive element~$\xi\in\FF_q^*$ and define the evaluation map
\begin{align*}
e_u \colon \FF_q^* \times \FF_q^* & \longrightarrow \FF_q \\
(\xi^i, \xi^j) & \longmapsto (\xi^i)^{a} (\xi^j)^{b}
\end{align*}
Then $w_u=(e_u((\xi^i,\xi^j)) : 0\leq i,j\leq q-2)$. Now let $P\subset\ZZ^2\otimes\QQ$ be a convex lattice polygon, and suppose that~$P$ is contained in the $[0,q-2]^2$ square. The \emph{toric code~$C_P(\FF_q)$ associated with~$P$} is defined to be the subspace spanned by the vectors~$\{w_u : u\in P\cap\ZZ^2\}$. This has block length~$n=(q-1)^2$ and dimension~$k$ at most $|P\cap\ZZ^2|$.

Generalised toric codes, introduced by Little~\cite{little1}, are constructed in a similar way, but allowing for the possibility of omitting lattice points from~$P$. In principle this allows for the deletion of short vectors in~$C_P$, thus potentially increasing the minimum Hamming distance of the resulting code. Equivalently, let $V\subset[0,q-2]^2$ be a set of lattice points. The \emph{generalised toric code} $C_V(\FF_q)$ associated with~$V$ is defined to be the subspace spanned by the vectors~$\{w_u : u\in V\}$. Clearly when $V=P\cap\ZZ^2$, where $P=\mathrm{conv}(V)$ is the convex hull of~$V$, we have $C_P=C_V$ and the concepts of toric codes and generalised toric codes coincide.

\begin{Ex*}
We work over~$\FF_3$ with primitive element~$\xi = 2$ in $\FF_3^*$, and consider the region~$[0,1]^2$ consisting of four lattice points. Then
\begin{equation*}
e_{(a,b)}((\xi^i,\xi^j))=(\xi^i)^a(\xi^j)^b=\xi^{ai+bj},\qquad\text{ where }(a,b)\in[0,1]^2\text{ and }(\xi^i,\xi^j)\in(\FF_3^*)^2.
\end{equation*}
In particular, a lattice point $u=(a,b)$ gives rise to the vector $w_u=(1,\xi^a,\xi^b,\xi^{a+b})$ (here we have implicitly fixed an ordering of the elements in $(\FF_3^*)^2$). Set~$V = \{(0,0),(0,1),(1,0)\}$. The generalised toric code~$C_V$ is the subspace spanned by the vectors $(1,1,1,1)$, $(1,1,2,2)$, and $(1,2,1,2)$ in $\FF_3^4$. Generator and parity check matrices (written in canonical form) for~$C_V$ are:
\begin{equation*}
 G = \begin{pmatrix}
 1 & 0 & 0 & 2 \\
 0 & 1 & 0 & 1 \\
 0 & 0 & 1 & 1
\end{pmatrix}
\qquad\text{ and }\qquad
P = \begin{pmatrix}
 1 & 2 & 2 & 1
 \end{pmatrix}
\end{equation*}
The minimum Hamming distance is readily seen to be two. Hence,~$C_V$ is a~$[4,3,2]_3$-code.
\end{Ex*}

Searches within the class of generalised toric codes have proven successful at finding champion codes~\cite{AmayaHarryVega,CarbonaraMurilloOrtiz,diego1,little1,kasprzyk1}. In particular, all generalised toric codes over~$\FF_q$, $q\leq 7$, and a subset over~$\FF_8$, were considered in~\cite{kasprzyk1}, identifying seven new champion linear codes. The method used there is exhaustive, and is computationally expensive due to the vast number of possible generalised toric codes, making a systematic search prohibitive when~$q\geq 8$. As one would expect, the calculation of the minimum Hamming distance~$d$ is a significant bottleneck in any search for new champion codes; for example, iIt is remarked in~\cite{kasprzyk1} that ``often these calculations would take millions of years if the code achieved its apparent minimum distance''.

\section{Machine learning experiments for generalised toric codes}\label{sec:appendixML}
\subsection*{A dataset of generalised toric codes over~$\FF_7$}
For each $5\leq m\leq 31$ we generated a dataset of $N=100,000$ distinct triples of the form~$(V, G, d)$, where~$V$ is a set of lattice points in $[0,5]^2$, $|V|=m$, chosen uniformly at random, $G$ is a generator matrix (in canonical form) for the corresponding code $C_V$, and~$d$ is the minimum Hamming distance of $C_V$; see Algorithm~\ref{alg:cap3}. We also experimented with including the parity check matrix~$P$ of~$C_V$ as a feature but, as expected given the relationship between~$G$ and~$P$, this was redundant.
\begin{algorithm}[t]
\small
\DontPrintSemicolon
\SetKwData{RandomPoints}{RandomPoints}
\SetKwData{GeneratorMatrix}{GeneratorMatrix}
\SetKwData{BZ}{BZ}
\KwData{A positive integer~$N$; a prime-power~$q$; a positive integer $m\leq (q-1)^2$.}
\KwResult{A set~$S^*$ of $N$ distinct triples of the form~$(V,G,d)$.}
\BlankLine
\emph{generate the set of sets of random lattice points}\;
$S\leftarrow\{\}$\;
\While{$|S|<N$}{
$S\leftarrow S\cup\{\RandomPoints(m)\}$\;
}
\BlankLine
\emph{compute the required data}\;
$S^*\leftarrow\{\}$\;
\For{$V\in S$}{
$C\leftarrow C_V(\FF_q)$\;
$G\leftarrow\GeneratorMatrix(C)$
\tcp*{calculate the canonical generator matrix}
$d\leftarrow\BZ(C)$
\tcp*{calculate the minimum Hamming distance}
$S^*\leftarrow S^*\cup\{(V,G,d)\}$\;
}
\caption{Generating datasets of size~$N$ over~$\FF_q$, for random lattice point sets of size~$m$.}\label{alg:cap3}
\end{algorithm}
Note that the number of subsets of lattice points of size~$m$ is~$\binom{(q-1)^2}{m}=\binom{36}{m}$. Thus, only in the range~$5 \leq m \leq 31$ do there exist at least 100,000 distinct sets of lattice points of size~$m$.

The dataset was generated using \texttt{Magma}~\cite{BosmaCannonPlayoust1997}, which can perform~BZ computations in parallel, on a computer cluster node with 96~Intel Xeon(R) Gold~6442Y processors (48 cores, 96 threads, 2.6-4.0 GHz). The calculation took approximately 24 hours. The resulting dataset, containing 2,700,000 examples, is illustrated in Figure~\ref{fig:F7data}. This dataset contains considerable variation in the frequency of the minimum Hamming distances, with the number of occurrences ranging from as few as two to over 600,000 instances.

Training on such an unbalanced dataset requires additional steps beyond simply partitioning the data into train and test sets randomly. Specifically, it is important to split each subset of the dataset with the same minimum Hamming distance into train and test sets separately, merging them into full train and test sets with re-shuffling afterwards. Class imbalance can be further mitigated through the application of oversampling/downsampling of the minority/majority classes. Model training for codes over~$\FF_7$ used cross-entropy loss, which allowed us to down-weight the losses for oversampled classes, and up-weight the losses for downsampled classes. The final dataset for~$\FF_7$ codes contained $\sim$550,000 elements, and was used for training with 9:1 train/test split. The subsets with less than 500 elements were oversampled with replacement to 500 elements, and the subsets with more than 50,000 examples were downsampled to 50,000. The intention was to obtain moderately imbalanced dataset with the smallest classes being no smaller than~$1\%$ of the largest class. The only class with less than 10 elements, corresponding to a minimum distance 13, was reserved for the test set only. 

\subsection*{A dataset of generalised toric codes over~$\FF_8$}
In the case of codes over~$\FF_8$, we were able to generate $N=100,000$ codes only for $4\leq m\leq 17$, following Algorithm~\ref{alg:cap3}. Due to the computation cost of the~BZ algorithm in higher dimensions, for $17\leq m\leq 39$ we reduced our target sample size to $N=1,000$, with an additional limit on the~BZ computation time of 1 minute. An additional 20,000 codes were generated for $m\in\{2,3, 40,\ldots,47\}$ afterwards.

The dataset was generated using \texttt{Magma} on a computer cluster node with 96~Intel Xeon(R) Gold~6442Y processors (48 cores, 96 threads, 2.6-4.0 GHz). The calculation took approximately 48 hours. The resulting dataset, containing 1,584,099 elements, is illustrated in Figure~\ref{fig:F8data}. The outlier at~$m=42$ is the result of a mistake during data collection, when 20,000 data points for~$m=41$ were added twice, and 20,000 points for~$m=42$ were erroneously discarded; this error was spotted only after the experiments were completed, and it was judged too computation expensive to repeat the processes, with no clear benefit in terms of the conclusions we could draw.. When~$m=2$ and~$3$ the sample size is smaller than 20,000 elements; a similar situation occurs when~$m = 47$. The variability in the number of examples by minimum Hamming distance in the region~$17\leq m\leq 40$ is caused by the limit on the allowed runtime for the~BZ algorithm, demonstrating the complexity of the procedure in this region. An unexpected feature of the~$\FF_8$ dataset is that dimension $k$ of each code is always equal to the number of lattice points $m$ used to generate the code -- this stands in contrast to the~$\FF_7$ dataset.

The minimum Hamming distance prediction task for the codes over~$\FF_8$ is expected to be more complicated than for codes over~$\FF_7$, but the amount of data we managed to collect was marginally smaller than for~$\FF_7$. To compensate for this limitation, we used a two-stage training procedure. A pre-training stage was done on a larger dataset with approximately 300,000 examples, which included subsets (codes with the same minimum Hamming distance) with more than 10,000 examples; subsets with more than 20,000 examples were downsampled to 20,000. For the training stage all the subsets were used, with down-sampling or over-sampling to 600. 

In goal was to train the model to extract useful features of codes on the common examples in the pre-training and then generalise this knowledge to all subsets in the training stage. In the later had a 10 times smaller learning rate to change the weights on a smaller scale than in pre-training to avoid losing any learned features.

The resulting performance is described in the section `Training and performance' (see~\S\ref{sec:results}). Here we will point out a couple of features found in the results. Over~$\FF_7$, Figure~\ref{fig:F7performance} reveals a notable trend: the model exhibits inferior performance for codes with minimum Hamming distances~$d = 2, 14, 22$, and~$27$. However, a worse performance can be explained by the scarcity of training examples only for~$d=7, 14$ and~$27$. In contrast, minimum Hamming distances~$d=2$ and~$22$ are well-represented in the dataset. These performance variations may indicate some intrinsic differences for codes with these minimum Hamming distances. Figure~\ref{fig:F8performance} reveals a similar pattern for codes over~$\FF_8$: the codes with minimum Hamming distances~$d=7, 14, 21$, and 28 are more difficult for the model to accurately predict. Again, this behaviour cannot be attributed to data scarcity. 

\subsection*{LSTM for codes over~$\FF_7$}
There are several architectures suitable for the sequence classification problem which we consider. 
We started our experiments with a smaller dataset, and used a long short-term memory~(LSTM) model~\cite{LSTM} to produce a baseline against which to compare future experiments.

\begin{figure}[t]
\centering
\includegraphics[width=1\textwidth]{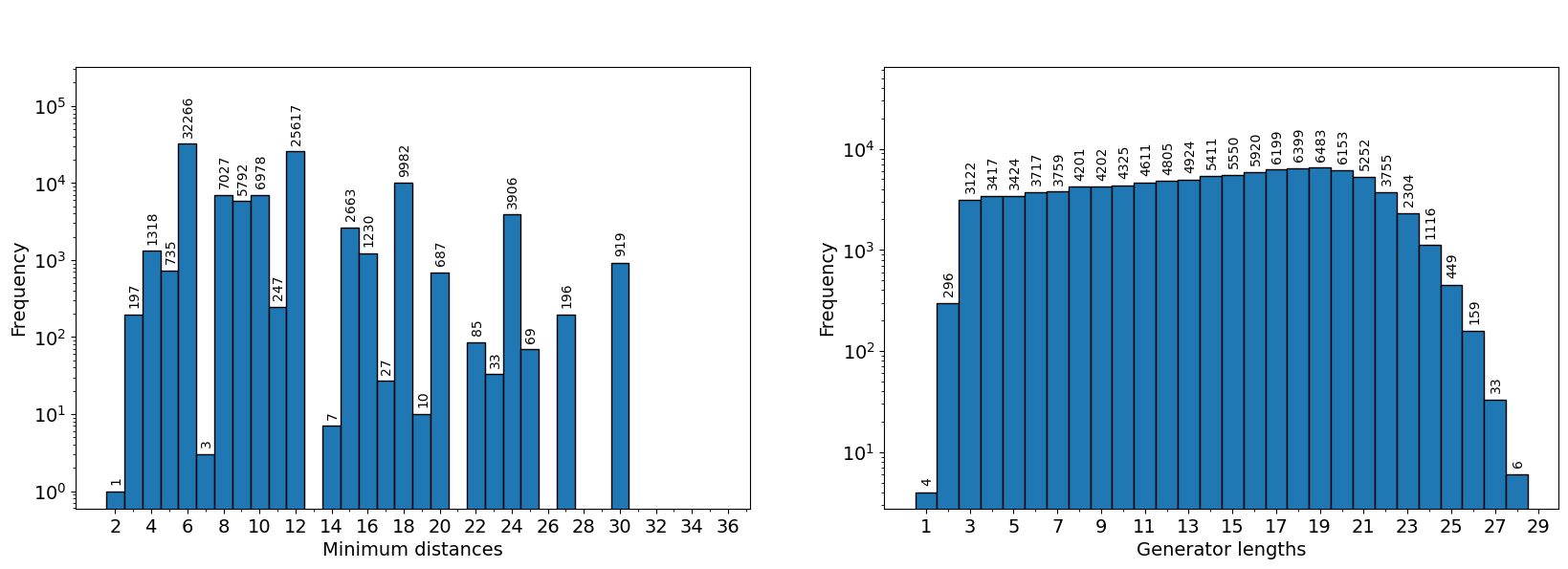}
\caption{Distributions of the minimum Hamming distance~$d$, and the dimensions~$k$, of generalised toric codes in the initial randomly generated dataset of 100,000 codes over~$\FF_7$, as discussed in the section `LSTM for codes over~$\FF_7$'.}
\label{fig:F7dataInit}
\end{figure}

A dataset of 100,000 triples~$(V, G, d)$ was generated by repeatedly choosing a random value for~$m$, $3\leq m \leq 28$, and then selecting $m$ random distinct lattice points for~$V$. This dataset is illustrated in Figure~\ref{fig:F7dataInit}. The result is an imbalanced dataset with many minimum Hamming distances not represented; this demonstrates the complexity of generating a representative dataset of generalised toric codes. For example, a rare minimum Hamming distance~$d=2$, which appears only once, originates from a generalised toric code with dimension~$k=22$. There were 3755 codes with dimension~$k=22$, and the majority ($88\%$) had minimum Hamming distance~$d=6$. Codes with dimension~$k=23$ may seem more likely to have minimum Hamming distance~$d=2$, but there were no such examples in the dataset.

This initial~$100,000$ codes dataset was randomly split into training and test sets in a 9:1 ratio. Because the dataset was very unbalanced and no measures to rebalance it were taken, the test set was missing minimum Hamming distances of 2, 7, 17, and 36. For input, the model takes batches of generator matrices. The target outputs~$d$ in the dataset were rescaled to lie in the range~$0\leq \widetilde{d}\leq1$ using the formula~$\widetilde{d} = (d - \mu)/\sigma$, where~$\mu = (30 + 3) / 2 = 16$ and~$\sigma = (30 - 3) / 2 = 13.5$. The model consisted of two layers of stacked~LSTMs. After the second layer, the hidden and cell states were extracted, concatenated, and passed to a linear layer projecting the output onto one continuous variable. The hyperparameters used are given in Table~\ref{tab:hyperparamsLSTM}.

\begin{table}[t]
\centering
\small
\begin{tabular}{llc}
\toprule
\textbf{Hyperparameter} & \textbf{Description} & \textbf{Value} \\
\midrule
\multirow{4}{*}{\begin{tabular}{@{}l@{}}Network\\architecture\end{tabular}} & Input dimension & 36 \\
 & Hidden dimension & 128 \\
 & Number of layers & 2 \\
 & Batch size & 4 \\
\midrule
\multirow{3}{*}{\begin{tabular}{@{}l@{}}Training\\parameters\end{tabular}} & Optimiser & AdamW \\
 & Learning rate & $1 \cdot 10^{-5}$ \\
 & Epochs & 200 \\
\bottomrule
\end{tabular}
\caption{The~LSTM architecture and training hyperparameters used for predicting minimum Hamming distances for generalised toric codes over~$\FF_7$.}
\label{tab:hyperparamsLSTM}
\end{table}

\begin{figure}[t]
\centering
 \includegraphics[width=1\textwidth]{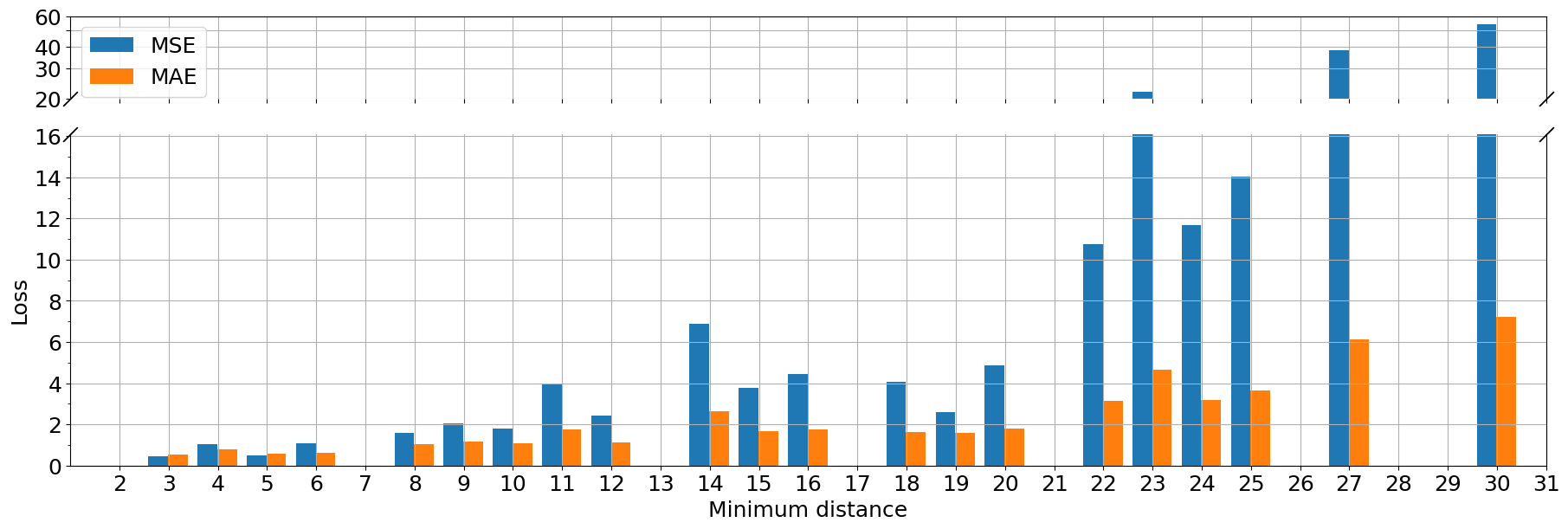}
 \caption{Losses on a test set of generalised toric codes over~$\FF_7$, collated by minimum Hamming distance, using an~LSTM model.}
 \label{fig:F7LSTMperformance}
\end{figure}

On the training set, the model’s~MAE was~$2.72$ and the accuracy within the tolerance~$d \pm 3$ was~$93.0\%$. On the test set, the model’s~MAE was~$3.02$ and the accuracy within the tolerance~$d \pm 3$ was~$91.2\%$. The per-minimum-distance metrics are illustrated in Figure~\ref{fig:F7LSTMperformance}. The model performed better for smaller minimum Hamming distances, and worse for larger distances. This model serves as a proof of concept that it is possible to learn to predict the minimum Hamming distance for a linear code to reasonable accuracy, and that the model is able to generalise to unseen data. 

Recall that the goal is to use the model as an estimator of the minimum Hamming distance, providing a fitness function for a genetic algorithm~(GA). During our experiments with~GAs, we observed that the final population after a~GA run was somewhat noisy: the range between the smallest and the largest minimum Hamming distance typically varied from 3 to 10 or more. As a consequence, the model does not need to be 100\% accurate in order to be efficient in this context; a~MAE error within a reasonable range would still be sufficient for practical use.

\subsection*{Toric transformer}
Recent advances in natural language processing~(NLP) are driven by the transformer architecture~\cite{vaswani2017attention}, which has demonstrated high effectiveness on various sequence processing tasks. As discussed in the section `Machine learning the minimum Hamming distance', the problem of predicting the minimum Hamming distance from the generator matrix can be viewed as a sequence classification task; it is therefore reasonable to attempt to modify the transformer to this setting. Compared to~NLP transformers, there are several architectural changes we need to make:
\begin{enumerate}
 \item input encoding of the field elements (for~$\FF_8$);
 \item masking to prevent padding from affecting results;
 \item summarisation along time to produce a sequence length independent output.
\end{enumerate}
We discuss these changes below when we describe the architecture in more detail. The general structure of the model is depicted in Figure~\ref{fig:TransformerScheme}, and is shared by both the model for codes over~$\FF_7$, and the model for codes over~$\FF_8$.

\subsection*{Input}
The input of the model is composed of a code tensor and a length vector. The code tensor is a \texttt{PyTorch}~\cite{pytorch} tensor containing a batch of generator matrices, padded with zeros to standardise the length. This tensor had three dimensions, denoted by $(B,T,C)$, where:
\begin{enumerate}
 \item $B$ is the~\emph{batch dimension}, i.e.\ the number of examples in a batch;
 \item $T$ is the~\emph{time dimension}, corresponding to the maximum number of rows in the generator matrices, i.e.~$(q-1)^2$ (for individual generator matrices, this dimension has a size equal to the dimension~$k$ of the linear code, but when padded and packed in the code tensor, it is equal to~$(q-1)^2$);
 \item $C$ is the~\emph{number of channels}, corresponding to the length of each row in the generator matrix, i.e.~$(q-1)^2$.
\end{enumerate}
The length vector stores the actual number of rows for each generator matrix in the batch. It can be used to distinguish between the rows of the generator matrix and any additional padding. 

\subsection*{Input embedding}
The input sequence -- the rows of the generator matrix in a batch -- is passed through an embedding layer. Embedding is a useful technique originating from~NLP. The idea is to map the elements of the input sequence into some higher-dimensional space. In~NLP, text data is first split into elementary building blocks of the sequence~(\emph{tokens}) which are then represented as vectors in higher dimensional vector space, allowing the~ML model to process them effectively. In our setting, the rows of the generator matrix are already in a vector form with entries of the vectors being elements of~$\FF_q$. Elements of~$\FF_7$ can naturally be regarded as integers in the set~$\{0,\ldots,6\}$, and thus the rows can be used as an input as is. This na\"{i}ve approach, however, does not work for~$\FF_8$. Instead fix a primitive element $\xi\in\FF^*_8$. Then the elements of~$\FF_8$ can be written as $\{0,1,\xi,\xi^2,\ldots,\xi^6\}$. Unfortunately these elements are not linearly independent, and although one-hot encoding was found to be efficient in the~$\FF_7$ case, it became unreasonable over~$\FF_8$. The alternative we used was to include an embedding layer with learnable representation; this is less efficient compared with a one-hot representation, in the sense that training to the same level of loss generally takes more epochs.

\subsection*{Attention blocks}
The attention mechanism in the transformer architecture and embedding align naturally with the structure of the problem of predicting minimum Hamming distances. After passing the rows of the generator matrix to the model, it will first represent them as vectors via the learned representation layer. These representations can then be compared in attention heads between each other, and useful information for the prediction of the minimum distance can be extracted. In our model, after the addition of positional encoding, the result was passed through an encoder-only set of self-attention blocks, each containing an attention layer and a feed-forward neural network.

\subsection*{Pooling layer}
In the standard transformer architecture, the number of dimensions of the input -- which in our model is of shape~$(B,T,C)$ -- is preserved at each stage. But for classification tasks, a single fixed-size vector per sequence is needed, and so we must eliminate the time dimension $T$ in the output. This is, we must introduce a pooling layer. There are many variants in the literature, with the simplest being: 
\begin{enumerate}
 \item \emph{BERT-like pooling.} Introduce the special classification token~\texttt{[CLS]} in the input, and then use the output at~\texttt{[CLS]} for classification.
 \item \emph{Average pooling.} Take the average of all output along the~$T$ dimension and use this for classification.
 \item \emph{Attention pooling.} Applying an attention mechanism to compute a weighted sum of sequence elements.
\end{enumerate}
It was shown in~\cite{tang2024pooling} that the later two are somewhat better than the first approach, and so we use them in our model. 

We used average pooling in the~$\FF_7$ case, and it demonstrated promising results despite being a very loose form of summarisation. The reason for this might be that within the attention layers, the rows of the generator matrix exchange information with each other and form more or less similar representations of the code. We expected the~$\FF_8$ case to be more complicated, and so attention pooling was applied. The idea here is to perform a transformation~$v\mapsto\mathrm{softmax}(v^T M) v$ with a learnable square matrix~$M$. This can reproduce the mean when~$M$ is proportional to the identity matrix and, therefore, should provide results that are no worse than the simple mean.

\subsection*{Output}
Next, a linear classification layer projecting from the embedding space to the output vector of dimension of~$(q-1)^2$ -- the number of possible minimum distances treated as~$(q-1)^2$ independent classes -- is applied to the pooled output. Finally, a softmax function was applied to obtain a vector~$\vec{p}$ of probabilities of classes. These probabilities were then used to compute the expectation of the minimum Hamming distance via~$\mathbb{E} [d] = \vec{p} \cdot \vec{i}$, where~$\vec{i}$ is the vector of class labels (i.e.\ target minimum Hamming distances). Since we were working with padded data, we also had to add masking to prevent padding elements from influencing the results in the attention blocks and pooling layers.

\subsection*{Loss functions}
Since we compute the (predicted) expected minimum Hamming distance, we could use the mean square error loss between the expected distance and the true distance. However, this loss would discard information about the individual probabilities of the classes. Instead, one should use either Wasserstein loss or cross-entropy loss. 

\emph{Wasserstein loss}~\cite{Wasserstein} has applications within Generative Adversarial Networks~(GANs). It measures the distance between probability distributions using some specified metric (we chose the Euclidean distance in experiments). We use this for codes over~$\FF_7$. \emph{Cross-entropy loss}~\cite{CrossEntropy} measures the difference between two probability distributions, but it is not a symmetric metric; it is more suitable for classification problems and thus has more use in~NLP. Our experiments demonstrated a superior performance for cross-entropy loss for codes over~$\FF_8$.

\subsection*{Hyperparameters}
Tables~\ref{tab:hyperparamsF7} and~\ref{tab:hyperparamsF8} list the hyperparameters of transformer models for generalised toric codes over~$\FF_7$ and~$\FF_8$, respectively.

\begin{table}[ht]
\centering
\small
\begin{tabular}{llc}
\toprule
\textbf{Hyperparameter} & \textbf{Description} & \textbf{Value} \\
\midrule
\multirow{8}{*}{\begin{tabular}{@{}l@{}}Network\\architecture\end{tabular}} & Embedding & No \\
 & Input dimension & 36 \\
 & Embedding dimension & 200\\
 & Attention heads & 10 \\
 & Transformer layers & 2 \\
 & Dropout & 0.2 \\
 & Loss function & Wasserstein-2 \\
 & Batch size & 32 \\
\midrule
\multirow{3}{*}{\begin{tabular}{@{}l@{}}Training\\parameters\end{tabular}} & Optimiser & AdamW\\
 & Learning Rate & $3 \cdot 10^{-4}$ \\
 & Epochs & 22\\
\bottomrule
\end{tabular}
\caption{The transformer architecture and training hyperparameters used for predicting minimum Hamming distances for generalised toric codes over~$\FF_7$.}
\label{tab:hyperparamsF7}
\end{table}

\begin{table}[ht]
\centering
\small
\begin{tabular}{llc}
\toprule
\textbf{Hyperparameter} & \textbf{Description} & \textbf{Value} \\
\midrule
\multirow{8}{*}{\begin{tabular}{@{}l@{}}Network\\architecture\end{tabular}} & Embedding & Learnable embedding \\
 & Input dimension & $49$ \\
 & Embedding dimension & 196\\
 & Attention heads & 7 \\
 & Transformer layers & 2 \\
 & Dropout & 0.2 \\
 & Loss function & Cross-entopy \\
 & Batch size & 128 \\
\midrule
\multirow{5}{*}{\begin{tabular}{@{}l@{}}Training\\parameters\end{tabular}} & Optimiser & AdamW\\
 & Learning Rate pre-trainig & $3 \cdot 10^{-5}$ \\
 & Learning Rate post-trainig & $3 \cdot 10^{-6}$ \\
 & Epochs pre-train & 120\\
 & Epochs post-train & 28\\
\bottomrule
\end{tabular}
\caption{The transformer architecture and training hyperparameters used for predicting minimum Hamming distances for generalised toric codes over~$\FF_8$.}
\label{tab:hyperparamsF8}
\end{table}

\subsection*{Improvements}
We used the models described above to search for champion codes using a~GA. We started our experiments with the model for generalised toric codes over~$\FF_7$, iteratively improving both the architecture and training procedure. These experiments showed that a number of modifications can be made to improve performance.
 
We found that instead of using the rows of the generator matrix as it is, it is more efficient to encode elements of rows using one-hot encoding. That is, we replaced entries in the input row by vectors of length seven containing zeros everywhere except one position unique to each of the integers $0,\ldots,6$ (and flattened afterwards). As a result, each input vector, that is, each row of the generator matrix, of length~$36$ is replaced by a vector of length~$252$. 

As described in the section `Datasets of generalised toric codes' above, we used an oversampling technique to rebalance the dataset. To avoid biases in the model's output distribution caused by duplicated examples, we down-weighted the oversampled examples during training. Using attention pooling instead of average pooling also demonstrated improved results.

Finally, in case of codes over~$\FF_7$, we used the entire dataset, which we partitioned into training and test sets. But unless we oversample some classes by a factor of thousands, the classes will be extremely unbalanced. To overcome this problem, we introduced a two-stage training procedure. First, in the pre-training stage, we used classes with many representatives (more than 10,000 in the case of codes over~$\FF_7$). This allows the model to learn general patterns useful for minimum Hamming distance prediction. Second, in the training stage, we aimed to have all available classes represented, and so we down-sampled classes with many representatives (down to 500 elements), and upsampled classes with too few representatives (up to 500 elements), and trained the pre-trained model on the resulting dataset. This second training stage uses a relatively small dataset (14,000 codes over~$\FF_7$) and so the model is prone to overfitting: early stopping should be used.

Overall, this showed excellent results for codes over~$\FF_7$; the next iteration of the model (not used with~GA) achieving cross-entropy loss of~0.61 on the pre-training dataset and~1.26 on the training dataset, and accuracy of~$84.8\%$ (within a tolerance of~$d \pm 3$) on the training dataset and~$90.3\%$ on the test set. The counter-intuitive higher accuracy on the test than than on the training set could potentially be explained by the fact that the training set has oversampled classes.

Applying this methodology to codes over~$\FF_8$ demonstrated less impressive results. Already at the pre-training stage, with plenty of data, the model was prone to overfitting after just~120 epochs. In general, there are approaches to overcoming this problem: increase the dataset size or decrease the number of parameters in the model. Knowing that the task of predicting minimum Hamming distance for generalised toric codes over~$\FF_8$ is complicated, we chose not to decrease the model size. One might hope to improve the results by increasing the size of the dataset, however data generation remains challenging due to the high computational cost of the~BZ algorithm.

\subsection*{Data availability}
Our datasets are available from GitHub~\cite{GAs}.
\subsection*{Code availability}
All code required to replicate the results in this paper is available from GitHub~\cite{GAs} under an MIT license.
\subsection*{Acknowledgements}
AK~is funded by EPSRC Fellowship~EP/N022513/1. QL~is funded by a Mitchell Fund award from Trinity College, Oxford.
\bibliographystyle{plain}
\bibliography{bibliography}
\end{document}